\documentclass[sigconf]{acmart}

\usepackage{amsmath}
\usepackage{multirow}
\usepackage[ruled,vlined]{algorithm2e}
\usepackage{tikz}
\usetikzlibrary{matrix, positioning, arrows.meta, calc, fit}

\newcommand{\vect}[1]{\ensuremath{\mathbf{#1}}}
\newcommand{\algo}[1]{\ensuremath{\mathbf{#1}}}
\newcommand{\mtx}[1]{\ensuremath{\mathbf{#1}}}

\DeclareMathOperator{\argmin}{argmin} 

\AtBeginDocument{%
  }

\setcopyright{acmlicensed}
\copyrightyear{2024}
\acmYear{2024}
\acmDOI{XXXXXXX.XXXXXXX}

\acmYear{YYYY}
\acmVolume{YYYY}
\acmNumber{X}
\acmDOI{XXXXXXX.XXXXXXX}
\acmISBN{}
\acmConference[Conference acronym 'XX]{Make sure to enter the correct
  conference title from your rights confirmation emai}{XXXX}{XXXX}

\begin{document}

\title[Probabilistic Linear Regression Attack on IronMask]{Hypersphere Secure Sketch Revisited: Probabilistic Linear Regression Attack on IronMask in Multiple Usage}


\author{Pengxu Zhu}
\affiliation{%
  \institution{Shanghai Jiao Tong University}
  \city{Shanghai}
  \country{China}}
\email{zhupengxu@sjtu.edu.com}

\author{Lei Wang}
\authornote{Corresponding author.}
\affiliation{%
  \institution{Shanghai Jiao Tong University}
  \city{Shanghai}
  \country{China}}
\email{wanglei\_hb@sjtu.edu.cn}

\renewcommand{\shortauthors}{P. Zhu and L. Wang}

\begin{abstract}

  Protection of biometric templates is a critical and urgent area of focus. \textbf{IronMask} demonstrates outstanding recognition performance while protecting facial templates against existing known attacks. In high-level, IronMask can be conceptualized as a fuzzy commitment scheme building on the hypersphere directly.
  We devise an attack on IronMask targeting on the security notion of renewability. Our attack, termed as \textbf{Probabilistic Linear Regression Attack}, utilizes the linearity of underlying used error correcting code. This attack is the first algorithm to successfully recover the original template when getting multiple protected templates in acceptable time and requirement of storage. We implement experiments on \textbf{IronMask} applied to protect \textbf{ArcFace} that well verify the validity of our attacks. Furthermore, we carry out experiments in noisy environments and confirm that our attacks are still applicable. Finally, we put forward two strategies to mitigate this type of attacks. 

\end{abstract}



\maketitle

\section{Introduction}

Biometric-based authentication has been under intensive and continuous investigation for decades. Recent works use deep neural networks to extract discriminative features from users' biometrics and achieve significant advances, such as facial images. \textbf{ArcFace}\cite{dengArcFaceAdditiveAngular2022a}, which is one of the state-of-the-art face recognition system, projects the face images to templates on a hypersphere and utilizes angular distance to distinguish identities. 

However, the exposure of facial templates has the potential to cause severe threats to both user privacy and the entire authentication system. There have been a great number of attacks to show the risky leakage of templates and some even can reconstruct the biometric images from corresponding templates, including face~\cite{maiReconstructionFaceImages2019,otroshishahrezaVulnerabilityStateoftheArtFace2024,shahrezaFaceReconstructionFacial2023}, fingerprint~\cite{cappelliFingerprintImageReconstruction2007}, iris~\cite{galballyIrisImageReconstruction2013} and finger vein~\cite{kaubaInverseBiometricsGenerating2021}. Therefore, the biometric template protection (BTP) technique is becoming pressing due to the security risks arising from widespread biometric-based authentication system.

BTP technique primarily achieves three goals: \textit{irreversibility}, \textit{renewability} and \textit{unlinkability} with considerable recognition performance. Irreversibility prevents the reconstruction of the original biometric templates, ensuring the security of the biometric template for one-time use. Renewability ensures the irreversibility with newly issued protected biometric template even though old ones have been leaked, enabling its multiple usage. Unlinkability guarantees that the protected biometric templates from the same person cannot be associated with a single identity. The unlinkability is the most robust property and considerably more challenging to achieve, compared to the irreversibility and the renewability.

The fuzzy-based scheme is a promising technique to implement BTP. Fuzzy-based schemes include fuzzy extractor~\cite{dodisFuzzyExtractorsHow2008}, fuzzy vault~\cite{juelsFuzzyVaultScheme2006} and fuzzy commitment~\cite{juelsFuzzyCommitmentScheme1999}. They commonly consist of two functions: \textit{information reconciliation} and \textit{privacy amplification}. Information reconciliation maps  similar readings to an identical value, while privacy amplification converts the value to an uniformly random secret string. Information reconciliation is frequently implemented by secure sketch based on an error-correcting code (ECC). Privacy amplification is accomplished by extractors or cryptographic hash function. The fuzzy-based scheme has been used to protect biometric templates in the binary space or set spaces with error-correcting codes~\cite{leeBiometricKeyBinding2007,arakalaFuzzyExtractorsMinutiaeBased2007a} and real-valued space $\mathbb R^n$ with lattice code~\cite{katsumataRevisitingFuzzySignatures2021,janaNeuralFuzzyExtractors2022}. Nonetheless, there have no fuzzy-based scheme on hypersphere without directly transforming to the binary space until the first result in~\cite{kimIronMaskModularArchitecture2021}. Kim et al. devise an error correcting code on hypersphere to build a secure sketch and in turn a fuzzy commitment scheme, named \textbf{IronMask}. They apply \text{IronMask} to protect facial template on \textbf{ArcFace} with template dimension as  $n=512$, recommending the error correcting parameter $\alpha = 16$. The combination achieves a true accept rate(TAR) of $99.79\%$ at a false accept rate(FAR) of $0.0005\%$ and providing at least $115$-bit security against known attacks. They claimed that their scheme satisfies irreversibility, renewability and unlinkability. To the best of our knowledge, it's the best BTP scheme to provide high security while preserving facial recognition performance without other secrets.

\subsection{Our Contributions.}

We analyze the renewability and the unlinkability of \textbf{IronMask}. With multiple protected templates, we devise an attack, named as \textbf{probabilistic linear regression attack}, that can successfully recover the original face template. Let $n$ denote the dimension of the output template and $\alpha$ denote the error correcting parameter. The algorithm's complexity is $O(n^p e^{c\alpha})$ where $p$ is relative to the underlying used linear regression solver algorithm and $c$ is relative to how many protected templates obtained. We apply the probabilistic linear regression attack on \textbf{IronMask} protecting \textbf{ArcFace} with $n=512$ and $\alpha=16$. The experiment is carried out on a single laptop with Intel Core i7-12700H running at 2.30 GHz and 64 GB RAM. The experimental results show that the execution time is around $5.3$ days when obtaining $n-1=511$ protected templates, and is around $621$ days when obtaining only $2$ protected templates for SVD-based linear solver. As for LSA-based linear solver, the execution time is around $4.8$ days when obtaining $3$ protected templates and around $7.1$ days when obtaining $281$ protected templates. We note that the attack algorithm is fully parallelizable, and hence using $m$ machines leads to a linear speedup of $m$ times. Moreover, we carry out experiments in noisy scenarios and demonstrate that our attacks are still applicable, substantiating its practical effectiveness in the real world. Furthermore, we propose two plausible defence strategies: Add Extra Noise in Sketching Step and Salting on strengthening \textbf{IronMask} in order to mitigate our attacks and reach around $63$-bit security.

\section{Related Works}

\subsection{Face Recognition on Hypersphere}
Face recognition employing deep neural networks typically involves two sequential stages. Firstly, the face image is ingested as input, and an embedded facial representation, serving as a template, is generated as output; Secondly, the similarity score between two face templates is computed, enabling the determination of whether two face images correspond to the same individual. 

Recent works have discovered that instead of using contrastive loss~\cite{chopraLearningSimilarityMetric2005} and triplet loss~\cite{schroffFaceNetUnifiedEmbedding2015a}, angular margin-based losses exhibits superior performance in training on large-scale dataset, such as SphereFace~\cite{liuSphereFaceDeepHypersphere2017}, \textbf{ArcFace}~\cite{dengArcFaceAdditiveAngular2022a} and MagFace~\cite{mengMagFaceUniversalRepresentation2021}. Under these face recognition frameworks, the face templates are constraint on a unit-hypersphere and the similarity score of two templates $\vect w_1, \vect w_2$ is calculated as cosine similarity $\mathrm{Score}(\vect w_1, \vect w_2) = \arccos{\frac{\vect w_1 \cdot \vect w_2}{|\vect w_1| |\vect w_2|}}$. 

\subsection{Fuzzy-based BTP on Face Recognition}
For neural network based face recognition systems, many protection techniques based on error-correcting code or secure sketch have been proposed. \cite{pandeyDeepSecureEncoding2016a,talrejaZeroShotDeepHashing2019} directly learn the mapping from face image to binary code, while \cite{aoInfraredFaceBased2009,mohanSignificantFeatureBased2019d} transform the face template from real-value to binary code by recording other information. However, they all suffer from great performance degradation due to the loss of discriminatory information during the translation from real space to binary space. Rathgeb et al. use LCSS(Linear Seprable Subcode) to extract binary representation of face template and apply a fuzzy vault scheme to protect it~\cite{rathgebDeepFaceFuzzy2022}. They claim around $32$ bits false accept security analysed in FERET dataset\cite{phillipsFERETDatabaseEvaluation1998}. Jiang et al. in \cite{jiangFacebasedAuthenticationUsing2022} also transform the face template to binary code but using computational secure sketch, which is based on DMSP assumption \cite{galbraithObfuscatedFuzzyHamming2019}, to implement face-based authentication scheme and achieve considerable performance. However the assumption is new and might require more analysis to strengthen its security. And the scheme lacks the security analysis of irreversibility and unlinkability. \textbf{IronMask}~\cite{kimIronMaskModularArchitecture2021}, which can be abstract as a fuzzy commitment scheme, directly builds protection scheme on hypersphere without translating to binary space and achieve slight performance loss than other protection techniques based on ECC. They demonstrate that their scheme satisfies irreversibility, renewability and unlinkability to known attacks in their parameter settings when protecting \textbf{ArcFace}. Under particular settings, they claim that it can provide at least $115$-bit security against known attacks.

\subsection{Attack on Fuzzy-based Scheme}

For secure sketch and fuzzy extractor on binary space, there have been analysis and attacks against the secure properties such as irreversibility, reusability and unlinkability. \cite{boyenReusableCryptographicFuzzy2004,simoensPrivacyWeaknessesBiometric2009} find that the original template can be recovered when getting multiple sketches from same template if the underlying error-correcting codes are different or biased. \cite{simoensPrivacyWeaknessesBiometric2009} finds an attack that can break the unlinkability of secure sketch. However, their attacks and analysis are focusing on the secure sketch within binary space and are not suitable for targeting the hypersphere. Until now, no efficient attack has been developed against the reusability and unlinkability of the secure sketch on the hypersphere. 

\section{Revisit HyperSphere ECC and Secure Sketch}

\subsection{Notations}

We denote the set $\{1, 2, \cdots, \rho\}$ as $[\rho]$. Denote general space/set, typically biometric template space, as math calligraphic such as $\mathcal{M}$. For particular space, we denote $\mathbb R^n$ as $n$-dimension real-value space and $S^{n-1}$ as hypersphere in $\mathbb R^n$. The vector in $\mathbb R^n$ is denoted by bold small letter such as $\vect v$ while the matrix is denoted by bold large letter such as $\mtx M$. We denote the set of orthogonal matrices in $\mathbb R^n$ as $O(n)$ without any ambiguity with respect to the notation for complexity. The angle distance between two vectors $\vect v, \vect w$ is defined as $Angle(\vect v, \vect w) = \arccos(\frac{\vect v \cdot \vect w}{|\vect v||\vect w|})$.

In this paper, we concentrate on the metric space $S^{n-1}$ with $Angle$ distance, as it is the embedding space of face template space $\mathcal M$ of \textbf{ArcFace}.

\subsection{HyperSphere Error Correct Code}

Here we recall the definition of the HyperSphere-ECC, preparing for constructing hypersphere secure sketch.

\begin{definition}[HyperSphere-ECC~\cite{kimIronMaskModularArchitecture2021a}]\label{def:hyper-ecc}
    A set of codewords $\mathcal C \subset S^{n-1}$ is called HyperSphere-ECC if it satisfies:
    \begin{enumerate}
        \item (Discriminative) $\forall \vect c_1, \vect c_2 \in \mathcal C$, $Angle(\vect c_1, \vect c_2) > \theta$;
        \item (Efficiently Decodable) There exists an efficient algorithm $\algo{Decode}$, such that $\forall \vect c \in \mathcal C, \forall \vect a \in S^{n-1}$, if $Angle(\vect a, \vect c) < \frac{\theta}{2}$, $\algo{Decode}(\vect a) = c$.
    \end{enumerate}
\end{definition}

In \cite{kimIronMaskModularArchitecture2021a}, Kim et al. devised a family of HyperSphere-ECC that can be efficiently sampled and decoded.

\begin{definition}\label{def:na-sc}\cite{kimIronMaskModularArchitecture2021a}
    For any positive integer $\alpha$, $\mathcal C_\alpha$ is a set of codewords which have exactly $\alpha$ non-zero entries. Each non-zero entries are either $-\frac{1}{\sqrt{\alpha}}$ or $\frac{1}{\sqrt{\alpha}}$. 
\end{definition}

\begin{theorem}\cite{kimIronMaskModularArchitecture2021a}
    The designed distance $\theta$ for $\mathcal C_\alpha$ is $\frac{1}{2}\arccos(1-\frac{1}{\alpha})$.
\end{theorem}

In real world, even for $\vect c \in \mathcal C, \vect a \in S^{n-1}, Angle(\vect a, \vect c) > \frac{\theta}{2}$, there are chances that $\algo{Decode}(\vect a ) = \vect c$. 

\begin{algorithm}[tbp]
    \caption{Sample and Decode Algorithms for HyperSphere-ECC $\mathcal C_\alpha$ in Definition \ref{def:na-sc}} 
    \SetKwProg{Fn}{Function}{}{}
    
    \Fn{\bf{Sample}($n, \alpha$) $\rightarrow \vect c$} {
        \KwData{dimension $n$, error parameter $\alpha$}
        \KwResult{$\vect c \in \mathcal C_\alpha$}
        Random choose $\alpha$ distinct positions $j_1, \ldots, j_\alpha$\;
        $\vect{c} \leftarrow (0, \ldots, 0)_n $\;
        \For{$i \in \{j_1, \ldots, j_\alpha\}$} {
            $c_i \xleftarrow{\$} \{-\frac{1}{\sqrt{\alpha}}, \frac{1}{\sqrt{\alpha}}\}$\;
        }
        Output \vect c\;
    }
    \Fn{\bf{Decode}($u, \alpha$) $\rightarrow \vect c$} {
        \KwData{$\vect u \in S^{n-1}$, error parameter $\alpha$}
        \KwResult{$\vect c \in \mathcal C_\alpha$}
        Find the best $\alpha$ positions $J = \{j_1, \ldots, j_\alpha\}$ such that
        $\forall j \in J, \forall k \in [n]/ J |u_j| \ge |u_k|$\;
        $\vect c \leftarrow (0, \ldots, 0)_n$\;
        \For{$i \in J=\{j_1, \ldots, j_\alpha\}$} {
            $c_i \leftarrow \frac{u_i}{|u_i|\sqrt{\alpha}}$\;
        }
        Output \vect c\;
    }
\end{algorithm}

\subsection{HyperSphere Secure Sketch and IronMask Scheme}

Secure sketch was first proposed in \cite{dodisFuzzyExtractorsHow2008}. It is a primitive that can precisely recover $\vect w$ from any $\vect w'$ close to $\vect w$ with public information while not revealing too much information of $\vect w$. It has been a basic component to construct fuzzy extractor~\cite{dodisFuzzyExtractorsHow2008} and fuzzy commitment~\cite{juelsFuzzyCommitmentScheme1999}. Here we recall the definition of the secure sketch.

\begin{definition}[Secure Sketch]
    An $(\mathcal M, t)$ secure sketch consists of a pair of algorithms $(\algo{SS}, \algo{Rec})$.
    \begin{itemize}
        \item The sketching algorithm \algo{SS} takes input $\vect w \in \mathcal M$, outputs sketch $s$ as public information.
        \item The recovery algorithm \algo{Rec} takes input $\vect w' \in \mathcal M$ and sketch $s$, outputs $w''$.
    \end{itemize}
    It satisfies the following properties:
    \begin{itemize}
        \item{Correctness:} if $dis(\vect w, \vect w')< t$, then $\algo{Rec}(\vect w', \algo{SS}(\vect w)) = \vect w$;
        \item{Security:} It requires that sketch $s$ does not leak too much information of $\vect w$, i.e. $\max_{\vect w}\Pr[\vect w|s=\algo{SS}(\vect w)] < \frac{1}{2^\lambda}$ in the sense of information view security. Or it's computational hard to retrieve $\vect w$ given sketch $s$, i.e. for any  Probabilistic Polynomial Time(PPT) Adversary $\mathcal A$, $\Pr[\mathcal A(s=\algo{SS}(\vect w)) = \vect w] < \frac{1}{2^\lambda}$ in the sense of computational view security.($\lambda$ is security parameter)
    \end{itemize}
\end{definition}

For space $\mathcal F^n$ with hamming distance, Dodis et al. proposes a general construction of secure sketch based on error-correcting code ~\cite{dodisFuzzyExtractorsHow2008}. They also devise a general construction on the transitive space $\mathcal M$, i.e. for any $a, b\in \mathcal M$, there exists an isometry transformation $\pi$ satisfying $b = \pi(a)$. Since hypersphere space is also transitive with orthogonal matrices, we can build a hypersphere secure sketch based on HyperSphere-ECC similar to error-correcting code.

\begin{definition}[HyperSphere Secure Sketch]\label{def:hyper-ss}
    Given a HyperSphere-ECC $\mathcal C$ with decode algorithm \algo{Decode} and design angle $\theta$, the hypersphere secure sketch can be constructed as below:
    \begin{itemize}
        \item Sketching algorithm \algo{SS}: on input $\vect w \in S^{n-1}$, $\vect c \xleftarrow{\$} \mathcal C$, randomly generate an orthogonal matrix $\mtx M$ that satisfies $\vect c = \mtx M\vect w$, output $\mtx M$ as sketch;
        \item Recovery algorithm \algo{Rec}: on input $\vect w' \in S^{n-1}$, orthogonal matrix $\mtx M$, compute $\vect v \leftarrow \mtx M\vect w'$, $\vect c' \leftarrow \mathbf{Decode}(\vect v)$, output $\vect w'' \leftarrow \mtx M^{-1}\vect c'$.
    \end{itemize} 

    It satisfies the following properties:

    \begin{itemize}
        \item Correctness: If $Angle(\vect w, \vect w') < \frac{\theta}{2}$, $Angle(\mtx M\vect w, \mtx M\vect w') < \frac{\theta}{2}$. Based on the correctness of \algo{Decode} algorithm, as $\mtx M\vect w = \vect c$, $\vect c' = \mathbf{Decode}(\mtx M\vect w') = \vect c$. Thus $\vect w'' = \mtx M^{-1} \vect c' = \mtx M^{-1} \vect c = \vect w$.
        \item Security: Since $\mtx M$ is randomized, if $\vect w$ is uniformly distributed on $S^{n-1}$, $\{\mtx M^{-1}\vect c| \forall \vect c\in \mathcal C\}$ is set of all possible inputs with equal probability of $\algo{SS}$. The probability for adversary guessing correct answer at one attempt is $\frac{1}{|\mathcal C|}$.
    \end{itemize}
\end{definition}

In \cite{kimIronMaskModularArchitecture2021a}, they implement an algorithm named \textbf{hidden matrix rotation} to generate the random orthogonal matrix $\mtx M$ with constraint $\vect c = \mtx M \vect w$.

\subsubsection{Tradeoff of Correctness and Security} For secure sketch based on ECC, the error correcting capability of ECC is an important parameter to control the usability and the security of the whole algorithm. To achieve high security, the error correcting capability would be sacrificed so as usability. In \cite{kimIronMaskModularArchitecture2021a}, they choose $\alpha = 16$ and $n= 512$ to achieve $|\mathcal C| = \binom{n}{\alpha} \cdot 2^{\alpha} \approx 2^{115}$ security with average degrades 0.18\% of true accept rate(TAR) at the same false accept rate(FAR) compared to facial recognition system without protection.

\subsubsection{Usage in Fuzzy Commitment and IronMask Scheme.} 

Secure sketch can be used in authentication combined with cryptographic hash function. The scheme is called fuzzy commitment\cite{juelsFuzzyCommitmentScheme1999}. The hash function is applied to secret codeword $\vect c$ to get $H(\vect c)$ which is stored in the server. To authenticate to the server, the user only needs to recover the codeword as $\vect c'$ by recovery algorithm of secure sketch, recompute $H(\vect c')$ and send it back to the server. And server checks whether $H(\vect c')$ and $H(\vect c)$ are equal. \textbf{IronMask}\cite{kimIronMaskModularArchitecture2021} utilizes this paradigm and replaces secure sketch by hypersphere secure sketch based on particular HyperSphere-ECC in Definition \ref{def:na-sc}. Hash function is computationally secure if the probability of correctly guessing codeword $\vect c$ in one trial is small. However, even for high probability of guessing secret codeword $\vect c$(e.g. $2^{-40}$ if $|\mathcal C| = 2^{40}$), take advantage of slow hashes, such as PBKDF2~\cite{moriartyPKCSPasswordBasedCryptography2017}, bcrypt~\cite{provosFutureAdaptablePasswordScheme} and scrypt~\cite{percivalScryptPasswordBasedKey2016}, it's still inapplicable to implement exhausted searching attack, offering realistic security.

\subsection{Threat Model}

Here we define a game version of multiple usage security/reusability of secure sketch. Note that if $t=0$, the multiple usage security degenerates to irreversibility.

\begin{definition}\label{def:mul-ss-security}
    Let $SS = (\algo{SS},\algo{Rec})$ be secure sketch's two algorithms. The experiment $\mathrm{SSMUL}_{\mathcal A, \theta, t}(n)$ is defined as follows:
    \begin{enumerate}
        \item The challenger $C$ chooses a biometric resource $\mathcal W$, samples $\vect w \in \mathcal W$ and sends $\algo{SS}(\vect w)$ to adversary $\mathcal A$;
        \item $\mathcal A$ asks $q\le t$ queries to challenger $\mathcal C$. $\mathcal C$ samples $\{\vect w_i\in \mathcal W, i=1,\ldots, q\}$ with constraints that $dis(\vect w, \vect w_i)\le\theta \wedge dis(\vect w_i, \vect w_j) \le \theta, \forall i,j \in \{1, \ldots, q\}$, calculates the response set $Q = \{\algo{SS}(\vect w_i), i=1,\ldots, q\}$ and sends $Q$ to $\mathcal A$;
        \item $\mathcal A$ outputs $\vect w'$. If $\vect w' = \vect w$, outputs $1$, else outputs $0$.  
    \end{enumerate}
    The secure sketch is secure with $(t+1)$ multiple usage if existing negligible function $negl$ such that $\Pr[\mathrm{SSMUL}_{\mathcal A, \theta, t}(n) = 1] < negl(n)$ for all \textbf{PPT} adversaries $\mathcal A$.
\end{definition}

The definition is very similar to the reusability of fuzzy extractor. However, our definition allows the attacker to control the distance of each sampled templates from same source while reusable fuzzy extractor does not \cite{aponEfficientReusableFuzzy2017,wenReusableFuzzyExtractor2018} or assumes too powerful attacker with ability to totally control shift distance between each templates in binary space \cite{boyenReusableCryptographicFuzzy2004}. We argue that our definition can more accurately capture the attacker's power to recover template in real scenarios. As in real world, the attacker is more probable to get multiple sketches from different servers but can not accurately control the shift distance enrolled each time.But it might get a vague quality report of each enrolled sketch and select the sketches that have similar qualities thus bounding the angle distance of pairs of corresponding unprotected templates, consistent with our security model.

\section{Probabilistic Linear Regression Attack}

\subsection{Core Idea}

Hypersphere secure sketch is secure from the information theoretical view in one-time usage. While if the same template $\vect w$ is sketched multiple times, the sketches can determine the original $\vect w$.

For example, if $\vect w$ is sketched twice, assume the sketches are $\mtx M_1, \mtx M_2$ and corresponding codewords are $\vect c_1, \vect c_2$. The codeword pair $(\vect c_1, \vect c_2)$ satisfies that $\vect c_2 = \mtx M_2 \mtx M_1^{-1} \vect c_1$. Since $\mtx M_1, \mtx M_2$ are random orthogonal matrix, $\mtx M=\mtx M_2\mtx M_1^{-1}$ can be seen as a random orthogonal matrix with only one constraint that maps $\vect c_1$ to $\vect c_2$. Sparsity of the codewords implies there are few other pairs of $(\vect c_1', \vect c_2')$ satisfying $\vect c_2' = \mtx M\vect c_1'$, so as corresponding template $\vect w$, otherwise $\mtx M$ should have other constraints and might even leak whole information of original template if $\mtx M\vect c \in \mathcal C_{\alpha} \forall \vect c \in \mathcal C_\alpha$(see in Section \ref{ch:limit_space}). By exhaustive searching in the space of codewords, the pair $(\vect c_1', \vect c_2')$ can be determined and the original template \vect w can be recovered as $\mtx M_1^{-1}\vect c_1'$. Even if the structure of HyperSphere-ECC can be used, it's possible to downgrade the computation complexity of recovering $\vect w$.

Based on the HyperSphere-ECC construction employed by \textbf{IronMask} in Definition \ref{def:na-sc}, there have the designed distance $\theta = \frac{1}{2}\arccos (1-\frac{1}{\alpha})$. For satisfactory accuracy, the codeword $\vect c$ should utilize a small value of $\alpha$(specifically, $n=512, \alpha=16$ are chosen). Therefore,  the codewords contain a preponderance of $(n-\alpha)$ zeros. From another view, given that matrix $\mtx M$  represents the output of the hypersphere secure sketch $\mathbf{SS}(\vect w)$ and relates to the codeword $\vect c$ via $\vect c = \mtx M \vect w$, it follows that $\vect w$ aligns orthogonally with a $\frac{n-\alpha}{n}$ fraction of the row vectors in $\mtx M$. It means that randomly selecting a row vector $\vect v'$ from $\mtx M$ yields a $\frac{n-\alpha}{n}$ probability that $\vect v'$ is orthogonal to $\vect w$.

In the realm of linear algebra, determining the vector $\vect w \in S^{n-1}$ necessitates a minimum of $(n-1)$ linear equations. Each sketch matrix $\mtx M$ derived from $\algo{SS}(\vect w)$ offers a $\frac{n-\alpha}{n}$ probability of correctly yielding a linear equation of the form $\vect w^T \vect v' = 0$. Therefore, if we possess $(n-1)$ sketches and randomly select a row vector from each, the probability of obtaining $(n-1)$ correct linear equations amounts to $(\frac{n-\alpha}{n})^{n-1}$, which approximates to $e^{\alpha}$ when $\alpha \ll n$. These equations are highly probably linear independent. Utilizing singular value decomposition(SVD), we can then deduce either the original vector \vect w(without noise) or a closely related vector $\vect w'$(with noise). Additionally, by utilizing the recovery algorithm of hypersphere secure sketch, we can reconstruct the original vector even when provided with a noisy solution $\vect w'$.

Furthermore, by fully exploiting HyperSphere-ECC inherent structure, we could reduce the number of required linear equations. Assuming that we possess $n$ sketches relating to $\vect w$, denoted as $\mtx M_1, \mtx M_2, \ldots, \mtx M_{n}$ with corresponding codewords $\vect c_1, \vect c_2, \ldots, \vect c_n$. We can deduce the equations
\begin{equation}
    c_{i} = \mtx M_i \mtx M_1^{-1} c_1, \forall 2\le i \le n
\end{equation}
. Defining $\mtx M'_i = \mtx M_i \mtx M_1^{-1}$, we can interpret $\mtx M'_i$ as sketches of $\vect c_1$. This allows us to solve for $\vect c_1$ using $(n-1)$ linear equations. However, given that the entries of $\vect c_1 \in \mathcal C_\alpha$ predominantly consist of zeroes and that the non-zero entries possess uniform norms, the task of determining $\vect c_1$ based on $k$ linear equations can be seen as the \textbf{Subset Sum Problem} or the \textbf{Sparse Linear Regression Problem}. Numerous algorithms exist for tackling such problems, such as \cite{lagariasSolvingLowdensitySubset1985,costerImprovedLowdensitySubset1992,schnorrLatticeBasisReduction} for the \textbf{Subset Sum Problem} and \cite{chenAtomicDecompositionBasis,candesStableSignalRecovery2006,caiOrthogonalMatchingPursuit2011,foucartMathematicalIntroductionCompressive2013,gamarnikSparseHighDimensionalLinear2019,gamarnikInferenceHighDimensionalLinear2021} for the \textbf{Sparse Linear Regression Problem}. We choose to use the Local Search Algorithm(LSA) introduced by Gamarnik and Zadik in \cite{gamarnikSparseHighDimensionalLinear2019} due to its effectiveness and simplicity of implementation. And other algorithms are more applicable when the coefficients of the linear equations are independent of the input vector($\vect c_1$), which does not align with our specific problem where the equation's value is zero.

\subsection{Details of Probabilistic Linear Regression Attack}\label{ch:overview-null-vector-attack}

The attack comprises three main components: the Linear Equation Sampler, the Linear Regression Solver, and the Threshold Determinant .

Initially, the linear equation sampler receives $t$ sketches, denoted as $\mtx M_1, \mtx M_2, \dots, \mtx M_t$, and sample rows from them to construct a single matrix $\mtx M$. Subsequently, the linear regression solver processes this matrix $\mtx M$ and strives to generate a solution vector $\vect w'$ satisfying $||\mtx M \vect w'|| \approx 0$. Finally, the threshold determinant  utilizes $\vect w'$ to recover candidate template $\vect w_r$ through recovery algorithm of secure sketch. This component then determines whether the recovered template is correct based on the predefined threshold $\theta_t$ and the angle between $\vect w_r$ and $\vect w_{r'}$, which is output of recovery algorithm with inputs of $\vect w_r$ and another sketch.

\subsubsection{Linear Equation Sampler}

\begin{algorithm}[tbp]
    \DontPrintSemicolon
    \caption{Linear Equation Sampler}\label{alg:line-sampler}
    \KwData{Matrices $\mtx M_1, \mtx M_2, \ldots, \mtx M_{t} \in \algo{SS}(\vect \cdot)$, type = "SVD" or "LSA",  $k$}
    \KwResult{$\mtx M$}
    \uIf{type = "SVD"} {
        $t' \leftarrow t$ \;
        \For{$i=1, \dots, t'$}{
            $\mtx M_i' \leftarrow \mtx M_i$ \;
        }
    }
    \ElseIf{type="LSA"} {
        $t' \leftarrow t-1$ \;
        \For{$i=1, \dots, t'$}{
            $\mtx M_i' \leftarrow \mtx M_{i+1}\mtx M_1^{-1}$ \;
        }
    }

    $l \leftarrow \lfloor k/t' \rfloor$ \;
    
    \For{$i=1, \dots, t'$} {
        $\{\vect v_{l{(i-1)} + j}^T | 1\le j \le l\}\gets$ Random select different $l$ row vectors of $\mtx M_i'$ \;
    }
    \For{$i=1, \dots, k-l*t'$} {
        $\{\vect v_{l*t' + i}^T\}\gets$ Random select row vector different from already sampled vectors of $\mtx M_i'$ \;
    }
    $\mtx M \gets$ vertical stack of $\vect v_1^T, \vect v_2^T, \ldots, \vect v_{k}^T$ \;
\end{algorithm}

The Linear Equation Sampler receives the sketches $\mtx M_1, \mtx M_2, \ldots, \mtx M_t$ derived from template $\vect w$ as input.  Depending on the linear regression solver's chosen algorithm, the sampler selects row vectors from these sketches differently. If the solver employs the SVD algorithm, the sampler randomly picks $k$ row vectors from the sketches. If solver uses LSA, the sampler first computes $t-1$ matrices $\mtx M_{i}' = \mtx M_{i+1} \mtx M_1^{-1} \forall 2\le i \le t$ and then randomly select $k$ row vectors from these matrices. Subsequently, the sampler vertically stacks the chosen vectors to form the matrix $\mtx M$ which is the input of the linear regression solver. The sampler's duty is to maximum the likelihood that $||\mtx M\vect w|| \approx 0$(or $||\mtx M \mtx M_1 \vect w|| \approx 0$).  We deem the matrix $\mtx M$ as "correct" if for each selected row vector, the entry in corresponding mapped codeword is 0, where $\vect w$ represents original input template. "Correct" matrix ensures that $||\mtx M \vect w|| \approx 0$.

\begin{definition}\label{def:correct-matrix}
    In Algorithm \ref{alg:line-sampler}, assume row vector $\vect v_i$ is sampled in $m$'th row of matrix $\mtx M'_j$ and the mapped codeword of $\mtx M'_j$ is $\vect c$, i.e. $\mtx M'_j \vect w = \vect c$ if type = "SVD" where $\mtx M_{j} = \algo{SS}(\vect w)$ or $\mtx M_{j+1} \vect w = \vect c$ if type = "LSA" where $\mtx M_{j+1} = \algo{SS}(\vect w)$. We say the row vector $\vect v_i$ sampled by linear equation sampler is "correct" if and only if the $m$'th entry of $\vect c$ is 0, i.e. $\vect c_{m} = 0$. We say the sampled matrix $\mtx M$ is "correct" if and only if all row vectors sampled are "correct". Otherwise, $\mtx M$ is "incorrect".
\end{definition}

\subsubsection{Linear Regression Solver}\label{ch:linear-svd-solver-optimized}

\begin{algorithm}[tbp]
    \DontPrintSemicolon
    \caption{Linear Regression Solver based on SVD}\label{alg:linear-solver-svd}
    \KwData{Matrix $\mtx M$ with size $k*n$ where $k \ge n-1$}
    \KwResult{$\vect w = \argmin_\vect w ||\mtx M \vect w||$ where $\vect w \in S^{n-1}$}
    $\vect w \leftarrow$ the eigenvector of matrix $\mtx M$ with smallest eigenvalue\;
\end{algorithm}

\begin{algorithm}[tbp]
    \DontPrintSemicolon
    \caption{Linear Regression Solver based on Local Search Algorithm(LSA)}\label{alg:linear-solver-lsa}
    \KwData{Matrix $\mtx M$ with size $k*n$, error correcting code parameter $\alpha$, hyper-parameter $d$, $t_{th}$}
    \KwResult{$\vect c$ or $\bot$}
    $t \leftarrow 0$\;
    \Repeat{$||\mtx M \vect c|| \le d$ or $t>t_{th}$} {
        $\vect c \leftarrow$ Random select codeword in $\mathcal C_\alpha$\;
        $t \leftarrow t + 1$\;
        \Repeat{$norm_{new} = norm_{pre}$} {
            $norm_{pre} \leftarrow ||\mtx M \vect c||$\;
            $norm_{new} \leftarrow ||\mtx M \vect c||$\;
            \For{$\vect c' \in \mathcal C_\alpha$ where $||\vect c' - \vect c|| = \frac{\sqrt{2}}{\sqrt{\alpha}}$} {
                $norm_{tmp} = ||\mtx M \vect c'||$\;
                \If{$norm_{tmp} < norm_{new}$} {
                    $norm_{new} \leftarrow norm_{tmp}$, $\vect c_{new} \leftarrow \vect c'$\;
                }
            }
            $\vect c \leftarrow \vect c_{new}$\;
        }
    }
    \If{$||\mtx M \vect c|| \le d$} { Output $\vect c$\;
    }
    \Else{  Output $\bot$\;
    }
\end{algorithm}

The linear regression solver takes the output matrix $\mtx M$ as input, solves the following optimization problem:
\begin{equation}\label{eq:lrs-target}
    \argmin_\vect w ||\mtx M \vect w ||, \vect w \in S^{n-1}
\end{equation}
Two algorithms are employed : SVD(Singular Vector Decomposition) and LSA(Local Search Algorithm). 

For SVD-based solver, a minimum of $(n-1)$ linear equations are required, and the output $\vect w'$ is a candidate solution for the original template. If only two sketches, $\mtx M_1, \mtx M_2$, are available in the step of linear equation sampler, an approximate solution of equation \ref{eq:lrs-target} can be obtained by solving a smaller matrix. When given 2 sketches, the task of linear equation sampler is equivalent to guessing $\frac{k}{2}$ zero entries in each corresponding codewords $\vect c_1 = (c_1^1, c_2^1, \dots, c_n^1), \vect c_2 = (c_1^2, c_2^2, \dots, c_n^2)$, thus total $k$ zero entries. Assuming $U_1$ and $U_2$ are the guessed set of zero-value entries in $\vect c_1$ and $\vect c_2$ respectively, and given that $\vect c_2 = \mtx M_2 \mtx M_1^{-1} \vect c_1$, we can formulate a set of linear equations. These equations relate the non-zero entries of $\vect c_1$ to the zero entries of $\vect c_2$ through the matrix $\mtx M_2 \mtx M_1^{-1}$. Therefore, assume $\mtx M_2 \mtx M_1^{-1} = (m_{ij})$, we have
\begin{equation}
    \sum_{j\notin U_1} m_{ij} c_j^1 = 0, \forall i \in U_2
\end{equation}
. By applying SVD, we could find a solution $\vect c'$ that minimizes the squared error of these equations, subject to the constraint that the entries of $\vect c'$ indexed by $U_1$ are zero.  The approximate solution for the original template is then given by $\vect w' = \mtx M_1^{-1} \vect c'$. This approach reduces the matrix size from the original $k * n$ to $\frac{k}{2} * (n - \frac{k}{2})$ by at least factor 2 when $k \ge n-1$.

For the LSA-based solver, the required number of linear equations exceeds $\alpha\log n$~\cite{gamarnikSparseHighDimensionalLinear2019}. The solution obtained is a codeword $\vect c' \in \mathcal C_\alpha$, and the candidate template solution is derived as $\mtx M_1^{-1} \vect c'$.

\subsubsection{Threshold Determinant}

The threshold determinant obtains the solution template vector $\vect w'$ from linear regression solver as an input and proceeds to attempt the recovery of the original template $\vect w$. Subsequently, it invokes the secure sketch's recovery algorithm utilizing $\vect w'$ and the sketch $\mtx M_1$ to obtain candidate template $\vect w_{r_1}$. Then invoke the the secure sketch's recovery algorithm utilizing $\vect w_{r_1}$ and the sketch $\mtx M_2$ to obtain another candidate template $\vect w_{r_2}$. Then the determinant calculates the angle $\theta'$ as $Angle(\vect w_{r_1}, \vect w_{r_2})$. If $\theta'$ surpasses a preset threshold $\theta_t$, the determinant returns a false output, indicating to the linear equation sampler that a new matrix should be generated for the linear regression solver to process. Otherwise, it outputs $\vect w_{r_1}$ as the recovered solution template. 

\begin{figure}[htbp]
    \centering
    \resizebox{\linewidth}{!}{
        \begin{tikzpicture}[every node=/.style={minimum size=2cm}]
\matrix(M1)[matrix of math nodes, left delimiter=(,right delimiter=)]
{
    v_{1,1} & v_{1,2} & \dots & v_{1,j} & \dots & v_{1,n} \\
    v_{2,1} & v_{2,2} & \dots & v_{2,j} & \dots & v_{2,n} \\
    \vdots  & \vdots  &  & \vdots  &  & \vdots\\
    v_{n,1} & v_{n,2} & \dots  & v_{n,j} & \dots & v_{n,n}\\
};

\node[fit=(M1-1-1)(M1-1-6), draw, blue, line width=2pt, inner sep=0pt, opacity=1.0] {};
\node[fit=(M1-2-1)(M1-2-6), draw, blue, line width=2pt, inner sep=0pt, opacity=1.0] {};
\node[fit=(M1-4-1)(M1-4-6), draw, red, dashed, line width=2pt, inner sep=0pt, opacity=1.0] {};

\matrix(Mv1)[left=1em of M1, matrix of math nodes]
{
    \mathbf{v_1} \\
    \mathbf{v_2} \\
    \vdots \\
    \mathbf{v_n} \\
};

\node[above=0 of M1](M1lb) {$M_1$};
\node[right=1em of M1](M1x) {$\times$};

\matrix(W1)[right=1em of M1x, matrix of math nodes, left delimiter=(,right delimiter=)]
{
    w_1 \\
    w_2 \\
    \vdots \\
    w_n \\
};

\node[above=0 of W1] {$\mathbf{w^t}$};

\node[right=1em of W1](M1d) {$=$};

\matrix(MW1)[right=1em of M1d, matrix of math nodes, left delimiter=(,right delimiter=)]
{
    0 \\
    0 \\
    \vdots \\
    1 \\
};

\matrix(M2)[below=2em of M1, matrix of math nodes, left delimiter=(,right delimiter=)]
{
    u_{1,1} & u_{1,2} & \dots & u_{1,j} & \dots & u_{1,n} \\
    u_{2,1} & u_{2,2} & \dots & u_{2,j} & \dots & u_{2,n} \\
    \vdots  & \vdots  &  & \vdots  &  & \vdots\\
    u_{n,1} & u_{n,2} & \dots  & u_{n,j} & \dots & u_{n,n}\\
};

\node[fit=(M2-2-1)(M2-2-6), draw, blue, line width=2pt, inner sep=0pt, opacity=1.0] {};
\node[fit=(M2-1-1)(M2-1-6), draw, red, dashed, line width=2pt, inner sep=0pt, opacity=1.0] {};
\node[fit=(M2-4-1)(M2-4-6), draw, blue, line width=2pt, inner sep=0pt, opacity=1.0] {};

\matrix(Mv2)[left=1em of M2, matrix of math nodes]
{
    \mathbf{u_1} \\
    \mathbf{u_2} \\
    \vdots \\
    \mathbf{u_n} \\
};

\node[above=0 of M2] {$M_2$};
\node[right=1em of M2](M2x) {$\times$};

\matrix(W2)[right=1em of M2x, matrix of math nodes, left delimiter=(,right delimiter=)]
{
    w_1 \\
    w_2 \\
    \vdots \\
    w_n \\
};

\node[above=0 of W2] {$\mathbf{w^t}$};

\node[right=1em of W2](M2d) {$=$};

\matrix(MW2)[right=1em of M2d, matrix of math nodes, left delimiter=(,right delimiter=)]
{
    1 \\
    0 \\
    \vdots \\
    0 \\
};

\node[above right=1em of MW2](Doc1) {Ratio of 0s $= \frac{n-\alpha}{n}$};

\draw [arrows = {-Computer Modern Rightarrow[line cap=round]}, thick] ($(MW1.east) + (1em, 0)$) -- (Doc1.north);

\draw [arrows = {-Computer Modern Rightarrow[line cap=round]}, thick] ($(MW2.east) + (1em, 0)$) -- (Doc1.south);

\node[fit=(M1)(Mv1)(M1lb)(M2)(Mv2)(Doc1), draw, inner sep=0.5cm] (Box1) {};

\newlength\boxwidth
\pgfextractx{\boxwidth}{\pgfpointdiff{\pgfpointanchor{Box1}{west}}{\pgfpointanchor{Box1}{east}}}

\node[below=5em of Box1] (Content) {
    \begin{tikzpicture}
        \matrix(M1)[matrix of math nodes, left delimiter=(,right delimiter=)]
        {
            \mathbf{v_{i^{1}_{1}}} \\
            \vdots  \\
            \mathbf{v_{i^{1}_\frac{n}{2}}}\\
            \mathbf{u_{i^{1}_{\frac{n}{2}+1}}} \\
            \vdots  \\
            \mathbf{u_{i^{1}_{n-1}}}\\
        };
        \node[above=1em of M1] {$M^1$};
        \node[right=1em of M1](M1dot) {$\dots$};
        \matrix(Mi)[right=1em of M1dot, matrix of math nodes, left delimiter=(,right delimiter=)]
        {
            \mathbf{v_{i^{k}_{1}}} \\
            \vdots  \\
            \mathbf{v_{i^{k}_\frac{n}{2}}}\\
            \mathbf{u_{i^{k}_{\frac{n}{2}+1}}} \\
            \vdots  \\
            \mathbf{u_{i^{k}_{n-1}}}\\
        };
        \node[above=1em of Mi] {$M^k$};
        \node[right=1em of Mi](Miequal) {$=$};
        \matrix(Mn)[right=1em of Miequal, matrix of math nodes, left delimiter=(,right delimiter=)_{(n-1)\times n}]
        {
            v_{i^{k}_{1},1} & v_{i^{k}_{1},2} & \dots & v_{i^{k}_{1},j} & \dots & v_{i^{k}_{1},n} \\
            \vdots  & \vdots  &  & \vdots  &  & \vdots\\
            v_{i^{k}_\frac{n}{2},1} & v_{i^{k}_\frac{n}{2},2} & \dots  & v_{i^{k}_\frac{n}{2},j} & \dots & v_{i^{k}_\frac{n}{2},n}\\
            u_{i^{k}_{\frac{n}{2}+1},1} & u_{i^{k}_{\frac{n}{2}+1},2} & \dots & u_{i^{k}_{\frac{n}{2}+1},j} & \dots & u_{i^{k}_{\frac{n}{2}+1},n} \\
            \vdots  & \vdots  &  & \vdots  &  & \vdots\\
            u_{i^{k}_{n-1},1} & u_{i^{k}_{n-1},2} & \dots  & u_{i^{k}_{n-1},j} & \dots & u_{i^{k}_{n-1},n}\\
        };
        \node[fit=(Mn-1-1)(Mn-6-1)(Mn-6-6), draw, blue, line width=2pt, inner sep=0pt, opacity=1.0] {};
        \node[right=1em of Mn](M1dot) {$\dots$};

        \node[below=1em of Mn](Mntimes) {${\times}$};
        \matrix(W)[below=1em of Mntimes, matrix of math nodes, left delimiter=(,right delimiter=)]
        {
            w_1 & w_2 & \dots & w_j & \dots & & w_n \\
        };
        \node[left=1em of W](Wlabel) {$\mathbf{w}$};
        \node[right=1em of W](Wequal) {$=$};
        \node[right=1em of Wequal] {$\mathbf{0}$};

        \draw [->] ([xshift=-1em]Wlabel.west) .. controls +(left:1cm) and +(down:1cm) .. ([xshift=0em, yshift=-1em]Mi.south) node[midway, sloped, below] {Orthogonal};;
    \end{tikzpicture}
};

\node[fit=(Content), draw, inner sep=0.5cm, minimum width=\boxwidth] (Box2) {};
\draw [arrows = {-Computer Modern Rightarrow[line cap=round]}, double, thick] (Box1.south) -- node[midway, fill=white, swap] {Sample $\frac{n}{2}$ rows on $M_1$ and $\frac{n}{2}-1$ rows on $M_2$} (Box2.north);

\end{tikzpicture}
    }
    \caption{Overview of probabilistic linear regression attack based on SVD on two matrices $\mtx M_1$ and $\mtx M_2$. The blue solid box indicates that the row vector is orthogonal to template $\vect w$ while the red dashed box is not. By randomly selecting $(n-1)$ row vectors, we finally get matrix $\mtx M^{k}$ that $\vect w$ is in null space of $\mtx M^{k}$. As $\mtx M^{k}$ is full of rank, the only one null vector is parallel to $\vect w$.}
\end{figure}

\subsection{Correctness and Complexity Analysis}

In this section, we present an analysis of the correctness and computational complexity of the probabilistic linear regression algorithm, specifically focusing on noiseless scenario. As for noisy environments, we demonstrate the practicality and efficiency of our algorithms through empirical experiments discussed in Section \ref{ch:experiments}.

\subsubsection{Correctness}

The proof of the correctness of the probabilistic linear regression attack comprises three primary steps. First, we demonstrate that the inverse probability($p_s$) of the output matrix $\mtx M$ from the linear equation sampler is "correct" in Definition \ref{def:correct-matrix}, which satisfies $\mtx M \vect w = 0$(or $\mtx M \mtx M_1 \vect w = 0$), is equal to $2^{\mathcal O(\alpha)}$. Secondly, we establish that if the input matrix is "correct", the solution derived from linear regression solver is parallel to original template $\vect w$. Lastly, we show that the threshold determinant effectively filters out solutions $\vect w'$ corresponding to "incorrect" sampled matrices.

Considering the linear equation sampler, let's assume the number of sampled rows is a multiple of the given matrices, i.e. $k = l * t'$($t'=t$ for SVD and $t'=t+1$ for LSA). The associated probability that $l$ sampled row vectors from each matrix are orthogonal to template $\vect w'$ is $\frac{\binom{l}{n-\alpha}}{\binom{l}{n}} \ge (1-\frac{\alpha}{n-l+1})^l$. Consequently, the probability $p_s$ of sampling all rows from $t'$ matrices exceeds $(1-\frac{\alpha}{n-l+1})^{l\times t'} = (1-\frac{\alpha}{n-l+1})^k$.  Given practical conditions where $\alpha \ll n$ and $k \le n-1$, we deduce that $p_s \ge (1-\frac{\alpha}{n-l+1})^k \ge (1-\frac{\alpha}{n-l+1})^{n-1} \approx e^{-\frac{\alpha n}{n-l+1}}$. Assume $t'$ is greater than $2$, we finally reach $s_p \ge e^{-2\alpha}$. Therefore, the inverse probability corresponds to $2^{\mathcal O(\alpha)}$.

Considering the linear regression solver, assume the solution vector of input matrix $\mtx M$ is $\vect w'$. Leveraging the correctness of SVD algorithm, we have $\mtx M(\vect w - \vect w') = 0$ if $\mtx M$ is "correct". For the SVD-based solver, we require the dimension of matrix $\mtx M$ is $(n-1) * n$ to ensure a rank of $(n - 1)$ with overwhelming probability. This guarantees that $\vect w$ is parallel to $\vect w'$. For the LSA-based solver, if $\mtx M$ is "correct", based on Theorem 2.7 from \cite{gamarnikSparseHighDimensionalLinear2019}, for a sufficiently large $k \ge \alpha\log(n)$ and small $\sigma$ as $n$ approaches infinity, we have the solution codeword $\vect c'$ and original codeword $\vect c$ satisfying $||\vect c - \vect c'|| \le \sigma$ and $\vect c = \vect c'$ for small enough $\sigma$. This implies that $\vect w' = \mtx M_1^{-1} \vect c'$ is parallel to $\mtx M_1^{-1}\vect c = \vect w$(noting that $-\mtx M_1\vect w$ is also a valid solution for $\argmin_\vect x||\mtx M \vect x||$). 

Lastly, considering the threshold determinant, if the matrix sampled from linear equation sampler is "correct", assume the solution vector of SVD-based solver is $\vect w'$(for LSA-based solver, solution vector is $\vect w' = \mtx M_i^{-1}\vect c'$). We observe that $\algo{DECODE}(\mtx M_i\vect w') = \algo{DECODE}(\vect w'^T \cdot \vect w\mtx M_i\vect w) = \vect w'^T \cdot \vect w\mtx M_i\vect w$ and $\mtx M_i^{-1}\algo{DECODE}(\mtx M_i\vect w') = \mtx M_i^{-1} \vect w'^T \cdot \vect w\mtx M_i\vect w = \vect w'$. Thus, we have 
\begin{align}
    \vect w_{r_1} &= \mtx M_1^{-1}\algo{DECODE}(\mtx M_1\vect w') = \vect w' \\
    \vect w_{r_2} &= \mtx M_2^{-1}\algo{DECODE}(\mtx M_2\vect w_{r_1}) = \mtx M_2^{-1}\algo{DECODE}(\mtx M_2\vect w') = \vect w'
\end{align}
, ensuring a zero angle between $w_{r_1}$ and $w_{r_2}$. Conversely, if the sampled matrix $\mtx M$ from linear equation sampler is "incorrect", the solution $\vect w'$ of linear regression solver should deviate significantly from the original template, making the candidate template $\vect w_{r_1}$ deviated from original template(both $\vect w$ and $-\vect w$). Thus $\vect w_{r_2} = \algo{Rec}(\vect w_{r_1})$ should also be deviated from $\vect w_{r_1}$, otherwise we find another codeword pair $\vect c_1' = \mtx M_1 \vect w_{r_1}$ and $\vect c_2' = \mtx M_2 \vect w_{r_2}$ satisfying $\vect c_1' = \mtx M_2 \mtx M_1^{-1} \vect c_2'$, which is impossible under the scenario that there are no other constraints for $\mtx M_2$ and $\mtx M_1$ but $\vect c_2 = \mtx M_2 \mtx M_1^{-1} \vect c_1$ where $\mtx M_2^{-1}\vect c_2 = \mtx M_1^{-1} \vect c_1 = \vect w$. Therefore, we could facilitate its exclusion through appropriate angle threshold settings. 

\subsubsection{Complexity}

\begin{figure}[!htbp]
    \includegraphics[width=1.0\linewidth]{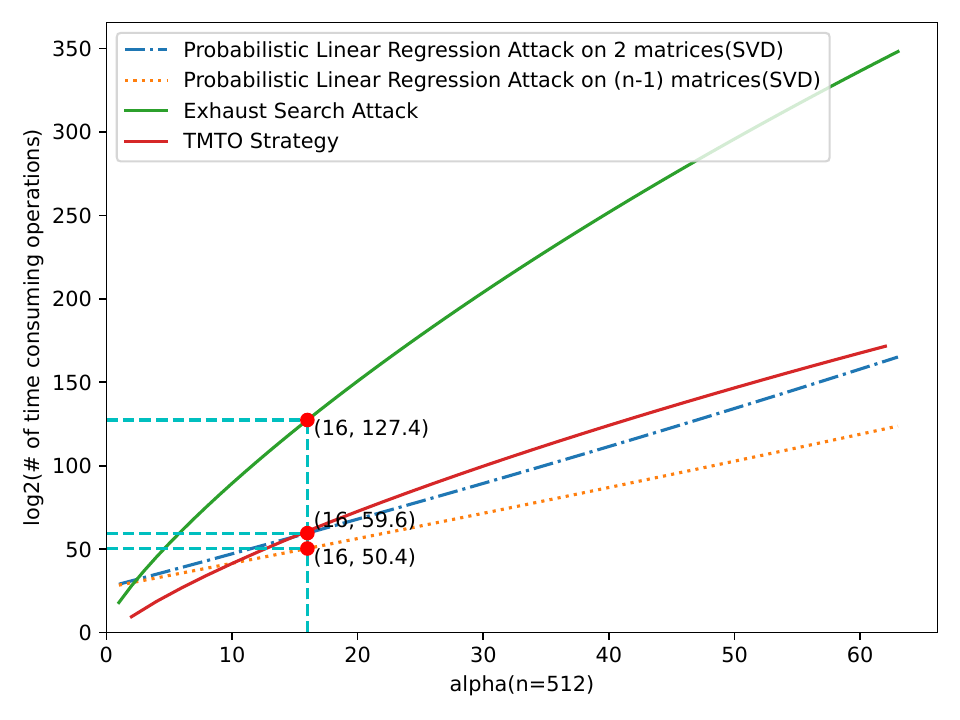}
    \centering
    \caption{$Log_2$ of the number of most time-consuming operations(complexity) of each algorithm according to different $\alpha$ with $n=512$. The complexity is $\mathcal O(n^3 e^{\alpha})$ for Algorithm \ref{alg:linear-solver-svd} and $\mathcal O(n^3 e^{2\alpha})$ for optimized SVD-based solver given $2$ sketch in Section \ref{ch:linear-svd-solver-optimized}. Here we take the constant number in the complexity of SVD algorithm as 1. However, for concrete algorithms, the constant number might be 8 or more. As this number is constant and small, we argue that it does not influence our conclusions.} \label{fig:complexity}
\end{figure}

In the context of the linear equation sampler, the inverse of the probability that the output matrix $\mtx M$ is "correct", which satisfies the condition $\mtx M \vect w = 0$(or $\mtx M \mtx M_1 \vect w = 0$), is given by $2^{\mathcal O(\alpha)}$, particularly $e^\alpha$ when $l$ equals to $1$ and $e^{2\alpha}$ when $l$ equals to $\frac{n}{2}$ where $k=n-1$ and $l = k / t'$. For the linear regression solver and threshold determinant components, the algorithms employed exhibit polynomial time complexity with respect to the matrix size $n$. Consequently, the overall time complexity of the entire algorithm can be expressed as $2^{\mathcal O(\alpha)} n^{b}$, where b is a constant representing the degree of the polynomial time complexity.

When it comes to the SVD linear regression solver, the SVD algorithm exhibits time complexity of $\mathcal O(n^3)$ and it operates on a sampled matrix of dimension $(n-1) \times n$. Considering the entire algorithm, the time complexity is $\mathcal O(e^{\alpha} n^3)$ for handling $n-1$ sketches, and $\mathcal O(e^{2\alpha} n^3)$ for handling 2 sketches.

Regarding the local search linear regression solver, each iteration carries a time complexity of $\mathcal O(n^2)$. The maximum iteration count is influenced by factors such as $\alpha, \sigma$ and $n$, at least $\alpha$ for random initial vector. Nevertheless, there is no explicit formula indicating the precise number of equations necessary to arrive at accurate solutions. Consequently, determining the overall algorithm's complexity based on the local search method remains elusive. However, through empirical observations in Section \ref{ch:experiments}, we hypothesize that the complexity of the local search-based algorithm is comparable to that of the SVD-based algorithm.

\subsection{Comparison with TMTO Strategy}

In \cite{kimIronMaskModularArchitecture2021a}, Kim et al. devise a time-memory-trade-off(TMTO) strategy to attack with two matrices, as solving $\vect{c_2} = \mtx M \vect{c_1}$ for $\mtx M = \mtx M_2\mtx M_1^{-1}$. The core idea is that each codeword $\vect c \in \mathcal C_\alpha$ can be seen as a combination of two codewords \vect{c'_1} and \vect{c''_1} from $C_{\frac{\alpha}{2}}$ with scalar $\frac{1}{\sqrt{2}}$ as $\vect{c_2} =\frac{1}{\sqrt{2}}\left( \mtx M \vect{c'_1} + \mtx M \vect{c''_1}\right)$. We only need to compute the smaller set $\{\mtx M\vect c'| c'\in\mathcal C_{\frac{\alpha}{2}}\}$ than exhaustive searching and check the pairs of $\vect c'$ and $\vect c''$ satisfied that the sum of $\mtx M\vect c'$ and $\mtx M\vect c''$ in particular entries are around $0, \pm \frac{\sqrt{2}}{\sqrt{\alpha}}$. By utilizing particular sort algorithms, the pairs that need to be checked can be greatly reduced so as the complexity. 

The obstacle of TMTO strategy is that it needs substantial storage. For particular settings $n=512, \alpha=16$, the required storage for storing codewords of $\mathcal C_8$ is at level EB. Considering the precision of the float number and noise in each sketching, to efficiently decrease the number of pairs to compare, the entries need to sum and sort would be more, making the storage requirement unacceptable. 

Due to the significant storage demands of the TMTO strategy, we chose not to implement it, focusing instead on providing a complexity analysis. As depicted in Figure \ref{fig:complexity}, when compared to the TMTO approach, our attack based on the SVD algorithm requires comparable computational resources when $\alpha \ge 16$($n=512$) given two sketches, and less computational resources when $\alpha \ge 16$($n=512$) given $n-1$ sketches if $\alpha < 64$. In specific scenarios ($n=512, \alpha=16$), the computational requirements under no noise of our algorithm are comparable to those of the TMTO strategy, with our algorithm requiring approximately $c*2^{60}$ multiplications versus $2^{60.8}$ additions of TMTO($c$ is a small constant relative to the SVD algorithm used). Moreover, our algorithm requires only a small amount of constant storage space, in contrast to the TMTO strategy, which demands large amounts of storage that are unacceptable in the proposed settings of IronMask. And the experiments in Section \ref{ch:experiments} demonstrate the effectiveness of our attacks while TMTO strategy might be not effective in same noise levels. Therefore, we contend that our algorithm is the first practical attack on IronMask in the real world. 

\subsection{Limit the Space of Secure Sketch}\label{ch:limit_space}

In Definition~\ref{def:hyper-ss}, the orthogonal matrix $\mtx M$ does not have other constraints so that there only few pairs $(\vect c_1, \vect c_2)$ satisfies $\vect c_2 = \mtx M_2 \mtx M_1^{-1} \vect c_1$. We may want that $\forall \vect w, \forall \mtx M_1, \mtx M_2 \in \algo{SS}(\vect w)$, $\forall \vect c\in \mathcal C$, $\mtx M_2 \mtx M_1^{-1} \vect c \in \mathcal C$ so that the utilization of multiple sketches does not result in any additional information leakage beyond that of a single sketch. Then our attacks will not work. It requires that $\mtx T = \mtx M_2 \mtx M_1^{-1}$ is not only an orthogonal matrix, but also maps $\mathcal C \rightarrow \mathcal C$. Here we give the format of matrices that maps $\mathcal C_\alpha \rightarrow \mathcal C_\alpha$(Proof seen in Appendix \ref{ch:attack-appendix2}).

\begin{theorem}\label{theorem:limit-space}
    The form of orthogonal matrices $\mathcal T = \{\mtx T\colon \mathcal C_\alpha \rightarrow \mathcal C_\alpha\}$ in $\mathbb R^{n}$ with constraints $\alpha \ne 2$ and $\alpha \ne n$ is:
    \begin{align}
        \begin{pmatrix}
            \pm \vect e_{i_1} & \pm \vect e_{i_2} & \cdots & \pm \vect e_{i_n}
        \end{pmatrix}
    \end{align}
    where $\vect e_{i_1}, \vect e_{i_2}, \cdots, \vect e_{i_n}$ is a permutation of unit vectors $\vect e_0, \vect e_1, \cdots, \vect e_n$.
\end{theorem}

The remaining problem is how to choose $\mtx M_1$ such that $\mtx T\mtx M_1$ does not reveal too much information of template $\vect w$. One strategy is to use naive isometry rotation\cite{kimIronMaskModularArchitecture2021} to define $\mtx M_1$ by fixing the mapped codeword. However, we find an attack that can retrieve the template with almost $60\%$ accuracy only given $\mtx T\mtx M_1$(see in Appendix \ref{ch:attack-appendix2}). The other strategy is to define $\mtx M_1$ with randomness of $\vect w$ and hash function $H$, i.e. $\mtx M_1 = H(\vect w)$. But the problem is that if $H$ is sensitive to the difference of $\vect w$, it's still vulnerable to our probabilistic linear regression attack in noisy scenario(see in Section \ref{ch:Noise}). Whether suitable orthogonal matrix given $\vect w$ without leaking too much information exists remains an open problem.

\section{Experiments}\label{ch:experiments}

We conduct experiments to attack \textbf{IronMask} protecting  \textbf{ArcFace} with specifically parameter settings($n=512, \alpha = 16$). The experiments are carried out on a single laptop with Intel Core i7-12700H running at 2.30 GHz and 64 GB RAM. For SVD-based linear regression solver, we use the function \algo{null\_space} of python library scipy\footnote{\url{https://scipy.org/}} and \algo{svd} of numpy\footnote{\url{https://numpy.org/}} library. For LSA-based linear regression solver, we implement using python and numpy\footnote{\url{https://numpy.org/}} library.

Let $r_k$ denote the expected number of matrices sampled by linear equation sampler until it samples the "correct" matrix $\mtx M$ in Definition \ref{def:correct-matrix}. Let $t_k$ denote the running time of linear regression solver that produces the solution of $\argmin_\vect w ||\mtx M\vect w||$ or terminates with a bot response. Define $p_k$ as the probability that the solution obtained from linear regression solver passes the threshold determinant when the sampled matrix is "correct". Then the expected running time $t_{all}$ of whole algorithm can be derived as $t_{all} = r_k * t_k / p_k$. As $r_k$ could be calculated by formula $r_k = \frac{1}{p_s}$ given $n, k, \alpha$, our task is to estimate $t_k$ and $p_k$ for varying number of equations $k$ under specific scenario. 

Note that LSA-based linear regression solver involves two iterations, and if the input matrix is "incorrect", the algorithm will reach the max number of outer iteration, denoted as $t_{th}$. Consequently, we have $t_k = t_{th} * t_{in}$ where $t_{in}$ is estimated running time of inner iteration. Assuming that the probability that LSA-based solver produces correct template in each outer iteration is $p_{out}$. The probability that LSA-based solver produces correct template given "correct" matrix before $t_{th}$ iterations is $1 - (1-p_{out})^{t_{th}}$. To minimize the overall running time is equal to minimizing $\frac{t_{th}}{1 - (1-p_{out})^{t_{th}}}$. Hence, we arrive at $t_{th} = argmin_{t} \frac{t}{1 - (1-p_{out})^t}$. Since $0< 1-p < 1$, we could deduce that $t_{th} = 1$ and thus $t_k = t_{in}$. 

\subsection{Experiments in Noiseless Scenario}

Since the expected iteration number $r_k$ that linear equation sampler outputs "correct" matrix $\mtx M$($||\mtx M\vect w|| \approx 0$) and the running time of solver are both influenced by the number of sampled linear equations($k$), we conduct experiments to determine the optimal setting with the shortest expected time for various values of $k$.

We conduct experiments on the noiseless scenario in each sketching step, specifically with $\theta=0$ in Definition \ref{def:mul-ss-security}. The estimated running time are shown in Table \ref{tab:expected-time}. Our experiments indicate that 
for SVD-based probabilistic linear regression solver, the expected running time is approximately 1.7 year given only 2 sketches and 5.3 day given $n-1=511$ sketches. As for the LSA-based probabilistic linear regression solver, the minimum expected running time is 4.8 day with 3 sketches and 7.1 day with 281 sketches. The results demonstrate that our algorithms are practical to attack \textbf{IronMask} applied to protect \textbf{ArcFace} in noiseless scenario. 

\begin{table*}[htbp]
    \caption{The estimated expected running time for successfully retrieving template $\vect w$ or $-\vect w$ with an expected value of 1 are analysed across different parameters $k$ with fixed $n=512$ for algorithms based on LSA-based solver in Algorithm \ref{alg:linear-solver-lsa} and SVD-based solver in Algorithm \ref{alg:linear-solver-svd}. Note that if the sketches are 2 for SVD-based solver, the algorithm is optimized mentioned in Section \ref{ch:linear-svd-solver-optimized}. $t_k$ is approximated by computing the mean of the running time over 1000 iterations. $p_k$ is estimated by calculating the proportion of successful times observed over maximum 20000 trials under the constraint that sampled matrix is "correct" in Definition \ref{def:correct-matrix}. } \label{tab:expected-time}
    \centering
    \begin{tabular}{cccccccc}
        \toprule
        Algorithm & \# sketches & $k$ & $r_k$ & Time($t_k$) & $p_k$ & $\theta_t$ & Time($t_{all}$) \\
        \midrule
        \multirow{2}{*}{SVD} & 2 & \multirow{2}{*}{511} & $2^{32.6}$ & 8.4 ms & 100\% & \multirow{2}{*}{$10^{\circ}$} & 1.7 year \\
        \cline{2-2}\cline{4-4}\cline{5-6} \cline{8-8}
        & 511 & & $2^{23.4}$ & 41ms & 100\% & & \color{red}{5.3 day} \\
        \hline

        \multirow{8}{*}{LSA} & \multirow{4}{*}{3} & 220 & $2^{11.35}$ & 102.0ms & $\frac{1}{1538.5}$ & \multirow{4}{*}{$10^{\circ}$} & \color{red}{4.8 day} \\
        \cline{3-3}\cline{4-6}\cline{8-8}

        & & 240 & $2^{12.54}$ & 108.6ms & $\frac{1}{833.3}$ & & 6.2 day \\
        \cline{3-3}\cline{4-6}\cline{8-8}

        & & 260 & $2^{13.75}$ & 116.0ms & $\frac{1}{512.8}$ & & 9.5 day \\
        \cline{3-3}\cline{4-6}\cline{8-8}

        & & 280 & $2^{15}$ & 130.0ms & $\frac{1}{219.8}$ & & 10.9 day \\
        \cline{2-8}

        & 261 & 260 & $2^{11.91}$ & 105ms
        & $\frac{1}{1666.7}$ & \multirow{4}{*}{$10^{\circ}$} & 7.8 day \\
        \cline{2-3}\cline{4-6}\cline{8-8}

        & 281 & 280 & $2^{12.83}$ & 114ms
        & $\frac{1}{740.7}$ & & \color{red}{7.1 day} \\
        \cline{2-3}\cline{4-6}\cline{8-8}

        & 301 & 300 & $2^{13.74}$ & 123ms
        & $\frac{1}{454.5}$ & & 8.86 day \\
        \cline{2-3}\cline{4-6}\cline{8-8}

        & 321 & 320 & $2^{14.65}$ & 105ms 
        & $\frac{1}{344.8}$ & & 10.8 day \\
        \bottomrule
    \end{tabular}
\end{table*}

\subsection{Experiments in Noisy Scenarios}\label{ch:Noise}

In real-world, it's better suited that the templates sketched each time have noise between each other, i.e. $Angle(\mtx M_i^{-1}\vect c_i, \mtx M_j^{-1}\vect c_j) <\theta'$ where $\vect c_i, \vect c_j$ are corresponding codewords to $\mtx M_i, \mtx M_j$. For example, in FEI dataset\cite{thomazNewRankingMethod2010}, the angle distance between different poses(p03, p04, p05, p06, p07, p08, p11, p12) of same face is below $36^{\circ}$ with $92\%$ probability and below $11^{\circ}$ with $0.3\%$ probability.

We argue that our algorithms possess the capability to accommodate medium value $\theta'$ with requirement of more iterations. Therefore, if the adversary have ability to choose sketches that maintain small angle distances within each corresponding original templates, they can still employ out attack to recover the original template.

However, upon consideration of noise, we discovered that even if the matrix is "correct", the threshold determinant alone cannot effectively discard solutions that deviate from the original template. This limitation arises because some candidate solutions $\vect w_{r_1}$ which are close to original template exhibit the characteristic that $\mtx M_2 \vect w_{r_1}$ are also close to the closest codeword, leading the algorithm to produce slight more candidate solutions. Nonetheless, there remains a high probability, denoted as $p_f$, that the output template of the algorithm is parallel to original template given a sampled "correct" matrix. Therefore, the expected running time $t_{all}$ that the algorithm finally output template $\vect w$ or $-\vect w$ is revised as $r_k * t_k / (p_k * p_f)$. The corresponding results are shown in Table \ref{tab:expected-time-noise}.

\begin{figure}[!htbp]
    \includegraphics[width=1.0\linewidth]{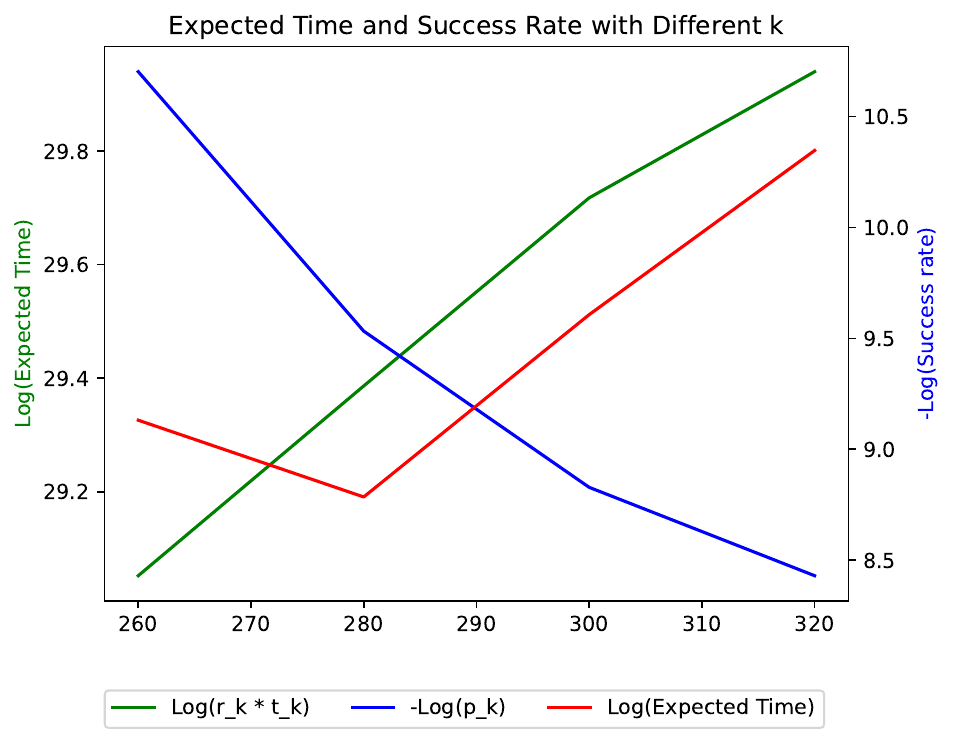}
    \centering
    \caption{$Log_2(r_k * t_k), Log_2(\frac{1}{ps}), Log_2(t_{all})$ of different $k$ using LSA-based attack algorithm when getting $k+1$ sketches in noiseless environments. The local minimum of $t_{all}$ is reached as $k\approx280$.} \label{fig:time-with-k}
\end{figure}

As $k$ increases,  $r_k \times t_k$ grows exponentially, while $\frac{1}{p_k \times p_f}$ decreases exponentially. Consequently, there exists a minimum 
$t_{all}$ in the mid-range of $k$, like the noiseless scenario in  Figure \ref{fig:time-with-k} with $p_f = 1$. Therefore, the Table \ref{tab:expected-time-noise} represents only the approximate minimum $t_{all}$ in noisy environments.

Table \ref{tab:expected-time-noise} reveals that our algorithms require a greater number of sampled linear equations, resulting in increased expected running time. Nevertheless, it is important to note that these algorithms are fully parallelizable. Therefore, by deploying additional machines or leveraging high-performance computing resources, we can effectively parallelize the algorithms, enabling our algorithm to recover the template within an acceptable running time, even when faced with noise levels of $36^{\circ}$.

\subsection{Experiments in Real-World Dataset}

\begin{table*}[htbp]
    \caption{Estimated expected running time using LSA-based probabilistic linear regression attacker on Real-World Dataset. $p_k * p_f$ is estimated by calculating the proportion of successful times observed over maximum 20000 trials under the constraint that sampled matrix is "correct" in Definition \ref{def:correct-matrix}.} \label{tab:real-world-dataset}
    \centering
    \begin{tabular}{cccccccc}
        \toprule
        Dataset & Noise($\theta'$) & $k$ & $r_k$ & Time($t_k$) & $p_k * p_f$ & $\theta_t$ & Time($t_{all}$) \\

        \hline
        FEI(p04, p05, p06) & $22.56^{\circ}$ & 300 & $2^{16.29}$ & 131.8ms & 
        $\frac{1}{847}$ & $40^{\circ}$ & 103.7 day \\
        
        \hline
        FEI(p03, p05, p08) & $28.56^{\circ}$ & 340 & $2^{18.97}$ & 129.0ms & 
        $\frac{1}{697}$ & $40^{\circ}$ & 1.47 year \\
        
        \bottomrule
    \end{tabular}
\end{table*}

Based on the previous section's findings, we have determined that the LSA-based linear regression solver exhibits superior performance. Therefore, we choose to employ the LSA-based attack algorithm, which necessitates the use of 3 sketches, for real-world simulations.

For our experiments, we selecte the FEI dataset, which utilizes the ArcFace neural network for feature extraction.  The FEI face database is a Brazilian face database that contains 14 images for each of 200 individuals, thus a total of 2800 images. We choose 2 sets of poses(p03, p05, p08 and p04, p05, p06) to simulate noisy environment and constrained environment. 

Table \ref{tab:real-world-dataset} shows that LSA-based probabilistic attack is applicable both in noisy environment and constrained environment, demonstrating the effectivness of our attacks in real world.

\begin{table*}[htbp]
    \caption{Estimated expected running time for different parameters $\theta_t, k$. $p_k * p_f$ is estimated by calculating the proportion of successful times observed over maximum 20000 trials under the constraint that sampled matrix is "correct" in Definition \ref{def:correct-matrix}.}\label{tab:expected-time-noise}
    \centering
    \begin{tabular}{ccccccccc}
        \hline 
        Noise($\theta'$) & Algorithm & \# sketches & $k$ & $r_k$ & Time($t_k$) & $p_k * p_f$ & $\theta_t$ & Time($t_{all}$) \\
        \midrule
        \multirow{4}{*}{$8.7^{\circ}$} & \multirow{2}{*}{SVD} & 2 & 522 & $2^{34.6}$ & 7.0 ms & 50\% & \multirow{2}{*}{$40^{\circ}$} & 6 year \\
        \cline{3-4}\cline{5-5}\cline{6-7} \cline{9-9}
        & & 531 & 531 & $2^{24.32}$ & 33.9ms & 78\% & & 10.5 day\\
        \cline{2-4}\cline{5-5}\cline{6-7} \cline{8-9}
        
        & \multirow{2}{*}{LSA} & 3 & 240 & $2^{12.54}$ & 110ms & $\frac{1}{800}$ & \multirow{2}{*}{$30^{\circ}$} & \textcolor{red}{6.1 day} \\
        \cline{3-4}\cline{5-5}\cline{6-7} \cline{9-9}

        & & 321 & 320 & $2^{14.65}$ & 112ms & $\frac{1}{350.9}$ & & 11.8 day \\
        
        \hline
        
        \multirow{4}{*}{$14^{\circ}$} & \multirow{2}{*}{SVD} & 2 & 532 & $2^{34.6}$ & 5.6 ms & 25\% & \multirow{2}{*}{$40^{\circ}$} & 18.9 year \\
        \cline{3-4}\cline{5-5}\cline{6-7} \cline{9-9}
        & & 551 & 551 & $2^{25.24}$ & 35.1ms & 40\% & & 40.2 day\\
        \cline{2-4}\cline{5-5}\cline{6-7} \cline{8-9}

        & \multirow{2}{*}{LSA} & 3 & 280 & $2^{15.01}$ & 126ms & $\frac{1}{392}$ & \multirow{2}{*}{$30^{\circ}$} & 18.8 day \\
        \cline{3-4}\cline{5-5}\cline{6-7} \cline{9-9}

        & & 321 & 320 & $2^{14.65}$ & 116ms & $\frac{1}{454.5}$ & & \textcolor{red}{15.8 day} \\
        
        \hline
        
        \multirow{4}{*}{$19^{\circ}$} & \multirow{2}{*}{SVD} & 2 & 552 & $2^{36.57}$ & 5.3 ms & 18\% & \multirow{2}{*}{$40^{\circ}$} & 95 year \\
        \cline{3-4}\cline{5-5}\cline{6-7} \cline{9-9}
        & & 591 & 591 & $2^{27.06}$ & 36.6ms & 26\% & & 229.6 day  \\
        \cline{2-4}\cline{5-5}\cline{6-7} \cline{8-9}

        & \multirow{2}{*}{LSA} & 3 & 280 & $2^{15.01}$ & 129ms & $\frac{1}{833.4}$ & \multirow{2}{*}{$30^{\circ}$} & 41 day \\
        \cline{3-4}\cline{5-5}\cline{6-7} \cline{9-9}

        & & 321 & 320 & $2^{14.65}$ & 115ms & $\frac{1}{1176.5}$ & & \textcolor{red}{40.5 day} \\
        \hline

        \multirow{4}{*}{$26^{\circ}$} & \multirow{2}{*}{SVD} & 2 & 592 & $2^{40.8}$ & 4.4 ms & 10\% & \multirow{2}{*}{$40^{\circ}$} & 2676 year \\
        \cline{3-4}\cline{5-5}\cline{6-7} \cline{9-9}
        & & 671 & 671 & $2^{30.73}$ & 38.5ms & 13.6\% & & 16 year \\
        \cline{2-4}\cline{5-5}\cline{6-7} \cline{8-9}

        & \multirow{2}{*}{LSA} & 3 & 320 & $2^{17.61}$ & 129ms & $\frac{1}{645.2}$ & \multirow{2}{*}{$40^{\circ}$} & \textcolor{red}{193 day} \\
        \cline{3-4}\cline{5-5}\cline{6-7} \cline{9-9}

        & & 381 & 380 & $2^{17.4}$ & 155ms & $\frac{1}{1111.1}$ & & 346 day \\
        
        \hline


        \multirow{3}{*}{$30^{\circ}$} & \multirow{1}{*}{SVD} & 771 & 771 & $2^{41.72}$ & 45.9 ms & 42.2\% & \multirow{1}{*}{$40^{\circ}$} & 147 year \\

        \cline{2-4}\cline{5-5}\cline{6-7} \cline{8-9}

        & \multirow{2}{*}{LSA} & 3 & 340 & $2^{18.97}$ & 132ms & $\frac{1}{869.6}$ & \multirow{2}{*}{$40^{\circ}$} & \textcolor{red}{1.87 year} \\
        \cline{3-4}\cline{5-5}\cline{6-7} \cline{9-9}

        & & 421 & 420 & $2^{19.2}$ & 155ms & $\frac{1}{1428.6}$ & & 4.34 year \\

        \hline


        \multirow{3}{*}{$36^{\circ}$} & \multirow{1}{*}{SVD} & 911 & 911 & $2^{45.3}$ & 49.5 ms & 13.4\% & \multirow{1}{*}{$45^{\circ}$} & $4.26 \times 10^4$ year \\

        \cline{2-4}\cline{5-5}\cline{6-7} \cline{8-9}

        & \multirow{2}{*}{LSA} & 3 & 380 & $2^{21.82}$ & 145ms & $\frac{1}{2857.1}$ & \multirow{2}{*}{$45^{\circ}$} & \textcolor{red}{48.65 year} \\
        \cline{3-4}\cline{5-5}\cline{6-7} \cline{9-9}

        & & 521 & 520 & $2^{23.82}$ & 178ms & $\frac{1}{2222.2}$ & & 185 year \\
        
        \hline


        \multirow{2}{*}{$43^{\circ}$} & \multirow{2}{*}{LSA} & 3 & 440 & $2^{21.82}$ & 168ms & $\frac{1}{10000}$ & \multirow{2}{*}{$50^{\circ}$} & \textcolor{red}{$4.8 \times 10^3$ year} \\
        \cline{3-4}\cline{5-5}\cline{6-7} \cline{9-9}

        & & 681 & 680 & $2^{23.82}$ & 225ms & $\frac{1}{4851.75}$ & & $8.2 \times 10^{4}$ year \\
        \hline
    \end{tabular}
\end{table*}
\section{Discussion of Plausible Defenses}

In the following content of this section, we discuss two strategies aimed at mitigating the impact of potential attacks against reusability. It is worth noting that these strategies are mutually independent, allowing us to combine them together to strengthen the hypersphere secure sketch against our attacks.

\subsection{Add Extra Noise in Sketching Step}\label{sec: add-extra-noise}

Our attacks and the TMTO strategy are effective primarily because the noise introduced between each sketching step is small, enabling us to identify a limited number of codeword pairs $(\vect c_1, \vect c_2)$ that meet the criterion of  $Angle(\mtx M_1^{-1}\vect c_1, \mtx M_2^{-1} \vect c_2)$ below a predefined threshold slightly larger than the estimated noise. However, if the noise between each sketching step becomes substantial enough to generate an excessive number of codeword pairs $(\vect c_1, \vect c_2)$ satisfying the same angle threshold criterion, our attacks become ineffective. A straight method to increase the angle between two templates is to introduce additional random noise to them.

Assume three unit vectors are $\vect w_1$, $\vect w_2$ and $\vect w_3$, take $\theta_{ij} = Angle(\vect w_i, \vect w_j)$, then we have $\vect w_3 = \cos \theta_{13} \vect w_1 + \sin \theta_{13} \vect u$ and $\vect w_2 = \cos \theta_{12} \vect w_1 + \sin \theta_{12} \vect v$ where $\vect u$ and $\vect v$ are all unit vectors and $\vect w_1^T\vect u  = 0$ and $\vect w_1^T \vect v = 0$. We have 
\begin{align*}
    \cos\theta_{23} &= \vect w_2^T \vect w_3 = (\cos \theta_{13} \vect w_1^T + \sin \theta_{13} \vect u^T) (\cos \theta_{12} \vect w_1 + \sin \theta_{12} \vect v) \\
    &= \cos \theta_{13} \cos \theta_{12} + \sin \theta_{13} \sin \theta_{12} \vect u^T \vect v
\end{align*}

If $\vect u$ or $\vect v$ is random, in expectation we have $\cos \theta_{23} \approx \cos \theta_{13} \cos \theta_{12}$. 

Thus assume the initial noise between two templates $\vect w_1$ and $\vect w_2$ are $\theta_i$, the random noise added in sketching step is $\theta_a$, and the extra-noisy versions of two templates are $\vect w_1'$ and $\vect w_2'$. We have $\cos Angle(\vect w_1', \vect w_2') \approx \cos \theta_i \cos^2 \theta_a$ and $\cos Angle(\vect w_1', \vect w_2) \approx \cos \theta_i \cos \theta_a$. 

For user, template $\vect w_2$ is utilized to retrieve template $\vect w_1'$. However, for an attacker, retrieve either $\vect w_1'$ or $\vect w_2'$ with corresponding sketches is necessary. The asymmetry in the noise encountered enables the user to still retrieve the template, while making it difficult for the attacker to carry out attacks due to the presence of larger noise..

\begin{figure}
    \centering
    \includegraphics[width=\linewidth]{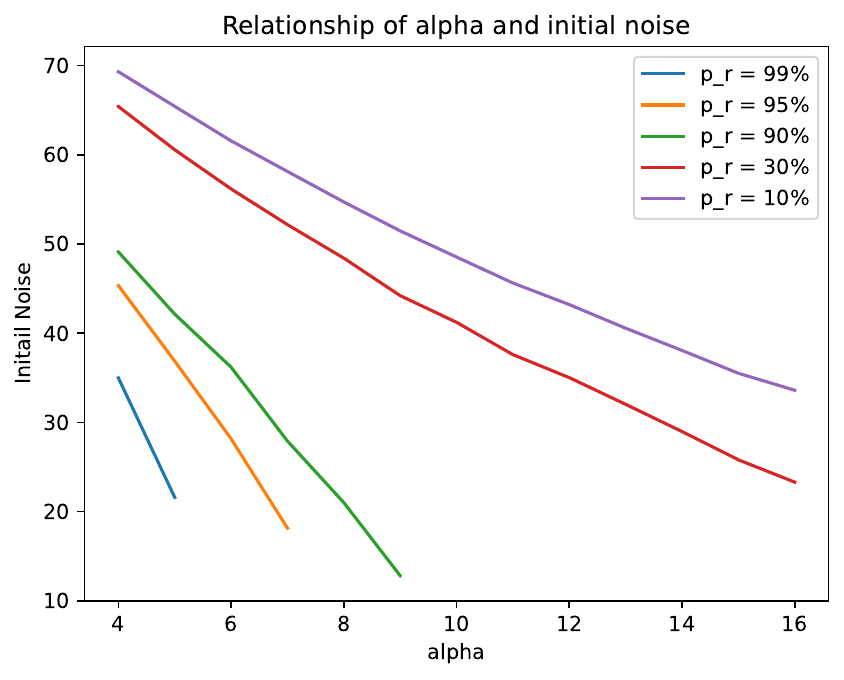}
    \caption{The relationship of initial noise $\theta_i$ and secure sketch parameter $\alpha(n=512)$ with different recovery probability $p_r$.}
    \label{fig:noise-trade-off}
\end{figure}

With a fixed template dimension of $n = 512$, we ensure that the distance between two templates, as perceived by an attacker, approximates the distance between a random unit vector and its nearest codeword, denoted as $\theta_r$. The relationship between the initial noise level and the success rate of the secure sketch recovery algorithm is in Table \ref{tab:noise-trade-off}. The table shows that as $\alpha$ decreases, while maintaining the same recovery success probability $p_r$, the initial noise $\theta_i$ increases. Therefore, by fixing $p_r$,  we can illustrate the relationship between $\theta_i$ and $\alpha$ in Figure \ref{fig:noise-trade-off}.

Assume an initial noise level of $36^\circ$, which is the criterion for the FEI dataset, to maintain a recovery probability of approximately $90\%$, we find that $\alpha\le 6$. This implies the brute-force space is reduced to less than $2^{50.46}$. However, if the attacker can obtain sketches from templates that are closer, the initial noise level should be lower than the criterion. For instance, in the FEI dataset, templates from poses p04, p05 and p06 are closer to each other than other poses. In such cases, the initial noise is $22.56^\circ$. To maintain the recovery probability high, we set $\alpha = 5$, achieving a recovery probability $p_r = 95\%$ in the FEI dataset with poses(p03, p04, p05, p06, p07, p08, p11, p12). Here, the size of the brute-force attack space is reduced from $2^{115}$ to $2^{43}$.

\begin{table}[htbp]
    \caption{The relation between initial noise($\theta_i)$ and success rate($p_r$) of recovery algorithm in secure sketch under different $\alpha$ with $n = 512$. As $\theta_r$ is expected noise in attackers' view which is equal to the distance between a random unit vector and its' nearest codeword, we have $\cos \theta_r = \cos \theta_i \cos^2 \theta_a$. } \label{tab:noise-trade-off}
    \centering
    \begin{tabular}{cccccc}
        \toprule
        $\alpha$ & $\theta_i$ & $\theta_a$ & $p_r$ & $\theta_r$ & $\log_2(|\mathcal C_\alpha|)$ \\

        \hline
        \multirow{3}{*}{16} & $0^\circ$ & $48.3^\circ$ & 56\% & \multirow{3}{*}{$63.8^\circ$} & \multirow{3}{*}{115} \\
        \cline{2-4}
         & $23.3^\circ$ & $46.1^\circ$ & 30\% & \\
         \cline{2-4}
         & $33.6^\circ$ & $43.2^\circ$ & 10\% & \\

        \hline
        \multirow{3}{*}{12} & $0^\circ$ & $50.9^\circ$ & 76\% & \multirow{3}{*}{$66.6^\circ$} & \multirow{3}{*}{91}  \\
        \cline{2-4}
         & $34.8^\circ$ & $45.9^\circ$ & 30\% & \\
         \cline{2-4}
         & $43.2^\circ$ & $42.3^\circ$ & 10\% & \\

         \hline
        \multirow{4}{*}{8} & $0^\circ$ & $54.3^\circ$ & 95\% & \multirow{4}{*}{$70.1^\circ$} & \multirow{4}{*}{64} \\
        \cline{2-4}
         & $21.0^\circ$ & $52.8^\circ$ & 90\% & \\
        \cline{2-4}
         & $48.4^\circ$ & $44.2^\circ$ & 30\% & \\
         \cline{2-4}
         & $54.7^\circ$ & $39.8^\circ$ & 10\% & \\

         \hline
        \multirow{4}{*}{4} & $0^\circ$ & $59.4^\circ$ & 100\% & \multirow{4}{*}{$75.0^\circ$} & \multirow{4}{*}{35} \\
        \cline{2-4}
         & $49.1^\circ$ & $51.1^\circ$ & 90\% & \\
        \cline{2-4}
         & $65.4^\circ$ & $38.2^\circ$ & 30\% & \\
         \cline{2-4}
         & $69.3^\circ$ & $31.2^\circ$ & 10\% & \\ 
        \bottomrule
    \end{tabular}
\end{table}

\subsection{Salting}

To carry out attacks targeting reusability, a minimum of two protected templates is necessary. By incorporating randomness into our protection algorithm, we can effectively slow down the attack algorithm's progress. In the context of the hypersphere secure sketch algorithm, we propose appending $n_{fake}$ additional random matrices, independent of the template $\vect w$, along with the output sketch matrix $\mtx M$. Subsequently, the recovery process is modified to invoke recovery algorithm with these $n_{fake} + 1$ matrices and take the recovery template closest to the input template as the final output.

From the attacker's perspective, acquiring two protected templates, each containing $n_{fake} + 1$ matrices, necessitates the examination of $(n_{fake} + 1)^2$ matrix pairs to employ the original attack algorithm. Consequently, we enhance the attack algorithm's complexity to a factor of $\mathcal O(n_{fake}^2)$ at the cost of $\mathcal O(n_{fake})$ times additional computations in recovery algorithm. In computational view,  this enhancement is equal to augmenting security by $\log_2(n_{fake})$ security bits.

For FEI dataset, if we take $\alpha = 5$ in Section \ref{sec: add-extra-noise}, the average runtime for recovery algorithm is $2.16\mu s$. To maintain the recovery time acceptable, we could take $n_{fake} = 2^{20}$. Then the time for recovery is approximate $2s$ while the time for brute-force attacker to primarily attack secure sketch is $2s * 2^{20} * 2^{43} /2 \approx 3.0 * 10^{12}$ year.

Due to the codeword space is shrinking to $2^{43}$, the fuzzy commitment scheme is not secure enough for directly solving $\vect c$ from $H(\vect c)$ where $\vect c \in \mathcal C_\alpha$. To enlarge the search space of hash function, we revise the commitment to be $H(\vect c, \mtx  M)$ to extend the search space from $\mathcal C_\alpha$ to $\mathcal C_\alpha \times \mathcal S_{\mtx  M}$ where $\mtx  M$ is the sketch of $\vect c$ and $\mathcal S_{\mtx  M}$ contains $\mtx M$ and $n_{fake}$ additional matrices. Thus for FEI dataset, the size of brute-force attacker's search space is $2^{20 + 43} = 2^{63}$. And if we set the time of hash function approximate $2$s, the complexity for directly attacking hash function is comparable to attacking secure sketch.

\section{Conclusion}
\textbf{IronMask} conceptualized as fuzzy commitment scheme is to protect the face template extracted by ArcFace in hypersphere, claims that it can provide at least 115-bit security against previous known attacks with great recognition performance. Targeting on renewability and unlinkability of \textbf{IronMask}, we proposed probabilistic linear regression attack that can successfully recover the original face template by exploiting the linearity of underlying used error correcting code. Under the recommended parameter settings on \textbf{IronMask} applied to protect \textbf{ArcFace}, our attacks are applicable in practical time verified by our experiments, even with the consistent noise level across biometric template extractions. To mitigate the impact of our attacks, we propose two plausible strategies for enhancing the hypersphere secure sketch scheme in \textbf{IronMask} at the cost of losses of security level and recovery success probabilities. To fully alleviate the error correcting code capability, future designs of hypersphere secure sketches and error-correcting codes should carefully consider the potential linearity of codewords, which may render them susceptible to attacks like the one we've presented.






\bibliographystyle{ACM-Reference-Format}
\bibliography{main.bib}

\appendix

\section{TMTO Strategy}

In \cite{kimIronMaskModularArchitecture2021a}, they describe a time-memory trade-off(TMTO) strategy to attack IronMask, here we revise the details of TMTO strategy to make it more applicable in real world's settings. 

Assume $\mtx M_1$ and $\mtx M_2$ are two sketches of biometric template $\vect w$, our target is to find codeword pair $(\vect c_1, \vect c_2)$ in $\mathcal C_\alpha$ such that $\vect c_2 = \mtx M \vect c_1$ for orthogonal matrix $\mtx M = \mtx M_2\mtx M_1^{-1}$. Let $\vect{c_1} = (c_{11}, c_{12}, \ldots, c_{1n})$, $\vect {c_2} = (c_{21}, c_{22}, \ldots, c_{2n})$ and $\mtx M = (m_{ij})$. The equation can be re-written as 
\begin{equation}\label{eq:TMTO-c2mc1-2}
    c_{11} m_{i1} + c_{12} m_{i2} + \cdots c_{1n} m_{in} = c_{2i}, \forall 1\le i \le n
\end{equation}
. As each codeword $\vect c \in \mathcal C_\alpha$ can be written as two components $\vect a, \vect b$ where each has exactly $\frac{\alpha}{2}$ non-zero elements and there is no positions that both $\vect a$ and \vect b are non-zero. We can rewrite Equation(\ref{eq:TMTO-c2mc1-2}) as
\begin{equation}\label{eq:TMTO-c2mc1-3}
    \sum_{\vect a = (a_1, \ldots, a_n) \in \mathcal C_{\frac{\alpha}{2}}} a_j m_{ij} + \sum_{\vect b = (b_1, \ldots, b_n) \in \mathcal C_{\frac{\alpha}{2}}} b_j m_{ij} = \sqrt{2}c_{2i}, \forall 1\le i \le n
\end{equation}
with constraint $\sqrt{2}\vect c_1 = \vect{a} + \vect b$. If we relax the constraint to that \vect a and \vect b have exactly $\frac{\alpha}{2}$ non-zero elements, the Equation(\ref{eq:TMTO-c2mc1-3}) can be simplified as
\begin{equation}\label{eq:TMTO-c2mc1-4}
    \sum_{\vect a \in \mathcal C_{\frac{\alpha}{2}}} a_j m_{ij} = \sqrt{2} c_{2i} - \sum_{\vect b \in \mathcal C_{\frac{\alpha}{2}}} b_j m_{ij} , \forall 1\le i \le n
\end{equation}
. As there are three elements $-\frac{1}{\sqrt{\alpha}}, 0, \frac{1}{\sqrt{\alpha}}$ for $c_{2i}$ and with high probability $c_{2i} = 0$ if $\alpha \ll n$, we can just assume $c_{2i} = 0$ for random selected $i$. Then we calculate all $t_{\vect a}^i = \sum_{\vect a \in \mathcal C_{\frac{\alpha}{2}}} a_j m_{ij}$ for some $i$, search them and find pairs that satisfies $t_{\vect a}^i + t_{\vect b}^i = 0$ for $\vect a, \vect b \in \mathcal C_{\frac{\alpha}{2}}$. As $t_{\vect a}^i = - t_{-\vect a}^i$, it only needs to find $\vect a, \vect b \in \mathcal C_{\frac{\alpha}{2}}$ such that $t_{\vect a}^i = t_{\vect b}^i$. The possible codeword $\vect c_1$ is equal to $\frac{1}{\sqrt{2}}(\vect a - \vect b)$.

In real world, since $t_\vect a$ is float number with limited precision and there's some noise in $c_{2i}$ in real settings, to make TMTO strategy work in these scenarios, we should calculate more $t_{\vect a}$ for different $i$'s and search by bucket with round-up.

The precise description of TMTO strategy attack is below:
\begin{enumerate}
    \item $\forall a \in \mathcal C_{\frac{\alpha}{2}}$, calculate the $t_{\vect a}^{i} = \sum a_j m_{ij}$ for chosen $i$'s;
    \item search $t_{\vect a}^{i}$ by bucket search or other search algorithms;
    \item find all different codewords $(\vect a, \vect b)$ that satisfies $t_{\vect a}^i = t_{\vect b}^i$ for all chosen $i$'s, check whether $t_{\vect a}^i = t_{\vect b}^i + x, \forall 1\le i\le n$ where $x = -\frac{\sqrt{2}}{\sqrt{\alpha}}, 0, \frac{\sqrt{2}}{\sqrt{\alpha}}$ and output $\frac{1}{\sqrt{2}}(\vect a - \vect b)$ as codeword $\vect c_1$.
\end{enumerate}

\noindent \textbf{Complexity}~ The probability that the equation $t^i_{\vect a} = t^i_{\vect b}$ is correct for random $i$ is $\frac{n-\alpha}{n}$. The number of codewords $\mathcal C_{\frac{\alpha}{2}}$ is $2^{\frac{\alpha}{2}}\binom{n}{\frac{\alpha}{2}}$. For each codeword in $\mathcal C_{\frac{\alpha}{2}}$, it needs $\frac{\alpha}{2}$ additions to compute $t^{i}$. Thus, with success expectation of $1$, we need total $\frac{\alpha n}{2(n - \alpha)} |\mathcal C_{\frac{\alpha}{2}}|$ additions and also need 
memory to store $|\mathcal C_{\frac{\alpha}{2}}|$ codewords with $t^i$. If the step 1-3 can be done together, we can early terminate the calculation of step 2 if satisfactory pairs of codewords have found. As for codeword $\vect c_{1} \in \mathcal C_{\alpha}$, there are $\binom{\alpha}{\frac{\alpha}{2}}$ pairs in $\mathcal C_{\frac{\alpha}{2}}$ that sums to $\sqrt{2}\vect c_1$. Therefore, the storage requirement and number of additions can be decreased by factor $\sqrt{\binom{\alpha}{\frac{\alpha}{2}}}$.

Here we assume that there are few pairs satisfying the conditions in step 3. However, since the precision of $t^i$ is limited, the satisfying pairs might be too large, making the computation cost of checking each pairs on other entries unacceptable. For example, if $t^i$ is 32-bit format float and distributed in range $(-1, 1)$, it can only exclude magnitude of $2^{33}$ pairs of codewords. Thus, to exclude enough pairs of codewords, we need to calculate $t^i$ for more different $i$'s. It will slightly enlarge the storage requirement with factor $\#i$(number of chosen $i$'s) and number of additions with factor $\#i * (\frac{n}{n - \alpha})^{\#i}$. If the the equation \ref{eq:TMTO-c2mc1-2} contains some noise, the required number of $i$'s would be more.

We ignore the complexity of search algorithm. As if the search algorithm is bucket search, the time and storage complexity is $O(|\mathcal C_{\frac{\alpha}{2}}|)$, comparable to the complexity of additions. 

For concrete settings as $n=512, \alpha=16$, the requirement of storage is around $2^{57.8}$ codewords and each codeword needs $8 * \log_2(512) = 72\text{ bit} = 9 \text{ bytes}$ with at least $4$ bytes for storage of $t^i$, which makes the storage larger than 2.8 EB. And the requirement of additions is around $2^{60.8}$.

\section{Remark on Limited Space of Secure Sketch}\label{ch:attack-appendix2}

Here we give an attack if the sketch is generated as in Section \ref{ch:limit_space}, i.e. $\algo{SS}(\vect w) = \mtx T \mtx M_1$ where $\mtx M_1$ is naive rotation matrix from $\vect w$ to predefined fixed codeword $\vect c_{fixed}$ and $\mtx T$ is defined in Theorem \ref{theorem:limit-space}. First, we give the proof of the Theorem \ref{theorem:limit-space}.

\begin{proof}[Theorem \ref{theorem:limit-space}]
    Assume $\mtx T = (t_{ij})$, $\vect a = (a_1, \ldots, a_n) \in \mathcal C_\alpha$ where $a_j = \frac{1}{\sqrt{\alpha}}$. Then $\vect a' = (a_1', \ldots, a_n') \in \mathcal C_\alpha$ where $\forall k\ne j$, $a_k'=a_k$ and $a_j'=-a_j$. Because of the definition of $\mtx T$, $\mtx T\vect a, \mtx T\vect a' \in \mathcal C_\alpha$. We have \begin{align}
        \mtx T\vect a - \mtx T\vect a' &= \vect c - \vect c' \\
        \frac{2}{\sqrt{\alpha}} t_{ij} &= \pm \frac{2}{\sqrt{\alpha}}, \pm \frac{1}{\sqrt{\alpha}}, 0 \\
        t_{ij} &= \pm 1, \pm 0.5, 0
    \end{align}
    $\forall i, j \in [n]$. Since $\mtx T$ is an orthogonal matrix, the norm of each row of $\mtx T$ is $1$. There are two cases for each row of $\mtx T$. One is that it consists of four positions filled with $\pm 0.5$, the other is that it only has one position filled with $\pm 1$. Here we prove that the first case is unsatisfactory by showing that if exist row $i$ of $\mtx T$ satisfies first case, $\exists \vect c \in \mathcal C_\alpha$, $\mtx T\vect c \notin \mathcal C_\alpha$.

    For row $i$ of $\mtx T$, assume $t_{ij_1}, t_{ij_2}, t_{ij_3}, t_{ij_4} = \pm 0.5$. If $\alpha = 1$, $\vect t_{i*} * \vect c = 0.5$ with $c_{j_1} = 1$. Then $\mtx T\vect c \notin \mathcal C_\alpha$. If $\alpha\ge 3$, construct $\vect c\in \mathcal C_\alpha$ so that $\forall 1\le k\le 3,c_{j_k} = sign(t_{ij_k}){\frac{1}{\sqrt{\alpha}}}$ and $c_{j_4} = 0$. Then $\vect t_{i*} * \vect c = \frac{3}{2\sqrt{\alpha}}$ and $\mtx T\vect c \notin C_\alpha$. 

    As each row of $\mtx T$ is equal to $\pm \vect e_i^T$ and $\mtx T$ is full of rank, the row vectors of $\mtx T$ can be seen as a permutation of $\vect e_0^T, \vect e_1^T, \cdots, \vect e_n^T$. The column vectors of $\mtx T$ are similar. So $\mtx T$ can be written as 
    \begin{align}
        \begin{pmatrix}
            \pm \vect e_{i_1} & \pm \vect e_{i_2} & \cdots & \pm \vect e_{i_n}
        \end{pmatrix}
    \end{align}
    where $\vect e_{i_1}, \vect e_{i_2}, \cdots, \vect e_{i_n}$ is a permutation of unit vectors $\vect e_0, \vect e_1, \cdots, \vect e_n$.
\end{proof}

\begin{algorithm}[!htbp]
    \DontPrintSemicolon
    \caption{Template Retrieve Attack on $\mtx M = \mtx{TR}$}\label{alg:rotation}
    \KwData{$\mtx M = \mtx{TR} \in O(n)$ with $\mtx T \in \mathcal T$ and $\mtx R$ is the native isometry rotation, threshold $\theta_t$, $m$}
    \KwResult{$\vect v$ or $\bot$}
    Create empty set $V$\;
    \For{$k = 2, \ldots, m$}{
        \For{$i' = 1, \ldots, n$ \tcp{repeat $n$ times} } {
            Random select distinct $k$ indices $I = (i_1 < \ldots < i_{k} \in [n])$\;
            Random select distinct $k$ indices $J = (j_1 < \ldots < j_{k}\in [n])$\;
            Select the submatrix $\mtx M'$ of $\mtx M$ with row indices $I$ and column indices $J$\;
            Compute the null vector $\vect u=(u_1, \ldots, u_{k})^T$ of submatrix $\mtx M'$\;
            Compute $\vect v = (v_1, \ldots, v_n)^T$ such that $v_{j_i} = u_i, \forall j_i \in J$ and $v_j = 0, \forall j \notin J$\;
            \If{$\mtx M\vect v$ contains exactly $i$ non-zero positions \tcp{make $\vect v$ satisfy $\mtx R \vect v = \vect v$}}{
                Store $\vect v$ in $V$\;
            }
        }
    }
    $\mtx M' \gets$ vertical stack of vectors in $V$\;
    $V' \gets$ approximate null vectors of $\mtx M'$\;
    \For{$\vect v \in V'$}{
        $\vect c' \gets \mathbf{Decode}(\mtx{M} \vect v)$\;
        \If{$Angle(\mtx{M} \vect v, \vect c') < \theta_t$} {
            Output $\vect v$
        }
    }
    Output $\bot$\;
\end{algorithm}

Then we recall the naive isometry rotation defined in \cite{kimIronMaskModularArchitecture2021}.

\begin{definition}[naive isometry rotation]
    Given vectors $\vect t, \vect c \in \mathbb R^n$, let $\vect w = \vect c - \vect t^T \vect c \vect t$ and $\mtx R_\theta = \begin{pmatrix}
        \cos \theta & - \sin \theta \\
        \sin \theta & \cos \theta
    \end{pmatrix}$ where $\theta = Angle(\vect t, \vect c)$. The naive rotation matrix $\mtx R$ mapping from $\vect t$ to $\vect c$ is:
    \begin{equation}
        \mtx R = \mtx I - \vect t \vect t^T - \vect w \vect w^T + \begin{pmatrix}
            \vect t & \vect w
        \end{pmatrix}\mtx R_{\theta} \begin{pmatrix}
            \vect t & \vect w
        \end{pmatrix}^T
    \end{equation}
    where $\mtx I$ is identity matrix.
\end{definition}

The naive isometry rotation $\mtx R$ can be seen as rotating $\vect t$ to $\vect c$ in the 2D plane $P$ extended by $\vect t$ and $\vect c$. Therefore, there are vectors in $(n-2)$ dimension subspace orthogonal to $P$ that satisfy $\mtx R \vect v = \vect v$. Restrict that $\vect v$ only has few non-zero positions, we have $\mtx T \mtx R \vect v = \mtx T \vect v$ filled with lots of $0$s. By guessing the non-zero positions in \vect v and zero positions in $\mtx T \mtx R\vect v$, we can calculate and filter to get $\vect v$. We can calculate the null space $P'$ of $\mtx M'$ made of these vectors. As $\vect v \perp P$, with enough vectors, we can greatly shrink the space $P'$ to $P$ and finally retrieve $P$. If $\vect t$ is the biometric template and $\vect c \in \mathcal C_\alpha$, it's easy to retrieve $\vect t$ knowing plane $P$. 

However, by experiments, we find that the first null vector $\vect v'$ of $\mtx M'$ is close enough to biometric template $\vect t$. Hence, we just take $\vect v'$ as possible candidate and use the sketch algorithm to try to retrieve original template $\vect t$ or $-\vect t$. The details are shown in Algorithm \ref{alg:rotation}.

With $n=512, \alpha=16$, Algorithm \ref{alg:rotation} can output original template $\vect w$ or $-\vect w$ with probability $\approx 60\%$ on 200 tests by setting $m=8$ and $\theta_t = 30^{\circ}$.

\end{document}